\documentclass[a4paper]{article}

\usepackage{stackengine}
\stackMath

\usepackage{enumitem}
\usepackage{rotating}
\newcommand{\invertediota}{\begin{sideways}\begin{sideways}$\iota$\end{sideways}\end{sideways}}

\usepackage{pdflscape}
\usepackage{mathabx}
\usepackage{philex}
\usepackage{latexsym}

\usepackage{amsthm}
\usepackage{bussproofs}
\usepackage{url}
\usepackage{pxfonts}
\usepackage{microtype}

\usepackage{natbib}

\newtheorem{theorem}{Theorem}
\newtheorem{lemma}{Lemma}
\newtheorem{definition}{Definition}
\newtheorem{corollary}{Corollary}

\begin{document}
\title{Normalisation for Negative Free Logics without and with Definite Descriptions}
\author{Nils K\"urbis}
\date{}
\maketitle

\begin{center}
Published in the \emph{Review of Symbolic Logic}
\url{https://doi.org/10.1017/S1755020324000157}\bigskip
\end{center}

\begin{abstract}
\noindent This paper proves normalisation theorems for intuitionist and classical negative free logic, without and with the $\invertediota$ operator for definite descriptions. Rules specific to free logic give rise to new kinds of maximal formulas additional to those familiar from standard intuitionist and classical logic. When $\invertediota$ is added it must be ensured that reduction procedures involving replacements of parameters by terms do not introduce new maximal formulas of higher degree than the ones removed. The problem is solved by a rule that permits restricting these terms in the rules for $\forall$, $\exists$ and $\invertediota$ to parameters or constants. A restricted subformula property for deductions in systems without $\invertediota$ is considered. It is improved upon by an alternative formalisation of free logic building on an idea of Ja\'skowski's. In the classical system the rules for $\invertediota$ require treatment known from normalisation for classical logic with $\lor$ or $\exists$. The philosophical significance of the results is also indicated. 
\end{abstract}

\section{Introduction}
It goes without saying that Russell’s theory of definite descriptions, expressions of the form `the $F$', carries great philosophical significance, and its importance in the development of analytic philosophy is hard to overstate. As there may not be a unique $F$, Russell proposed that `The $F$ is $G$' means `There is exactly one $F$ and it is $G$': the definite description disappears upon analysis and is not a genuine singular term.\footnote{See \citep{russellondenoting}, \citep[Ch. 16]{russellintromathphil} and \citep[Introduction, Ch. 3, Sec. (1), and Part I, Section B, $\ast$14]{russellwhitehead}.}

Despite its paradigmatic status, Russell's theory has not met with universal acceptance. A motive in the development of free logic by Hintikka, Lambert and others was the formalisation of alternative theories.\footnote{See \cite{bencivengahandbook} for an overview and the further references later in the text.} Free logic does not require that singular terms refer nor that domains of quantification be non-empty. Definite descriptions are treated as genuine singular terms. Nonetheless, negative free logic retains some Russellian spirit: atomic formulas containing singular terms cannot be true unless the terms refer. 

The present paper investigates systems of natural deduction for classical and intuitionist negative free logic with and without definite descriptions from a proof-theoretic perspective. Normalisation theorems for various systems formalised in Gentzen's natural deduction are proved. They establish that maximal formulas -- major premises of elimination rules that are concluded by an introduction rule -- can be removed from deductions by applying reduction procedures that transform them so that these inferences are avoided. Normalisation theorems are analogous to Gentzen's \emph{Hauptsatz} for sequent calculus.\footnote{Gentzen's Nachlass, edited by von Plato, also contains a normalisation result \citep{platocellar}.}

I shall begin with systems without definite descriptions familiar from the literature. To my knowledge, no normalisation theorems have been proved for them. To fill this gap is the first contribution of this paper. The proofs largely follow Prawitz's method \citep{prawitznaturaldeduction}. For the classical system with $\invertediota$ a method of Andou's is adapted \citep{andounormalisation}. Normalisation for free logic requires considering new cases of maximal formulas that do not occur in standard classical and intuitionist logic. Furthermore, I shall investigate the form of normal deductions and consider a suitable subformula property. Due to the rules for the quantifiers and those characteristic of negative free logic, this notion has to be rather restricted. Thus I shall propose an alternative way of formalising free logic inspired by Ja\'skowski that improves on the situation. 

The next contribution is to prove normalisation theorems for systems with the $\invertediota$ operator to formalise definite descriptions. Rules for $\invertediota$ suitable for negative free logic were given by Neil Tennant. Tennant also gives reduction procedures for maximal formulas of the form $\invertediota xF=t$, but to my knowledge, no one has explicitly proved normalisation for Tennant's systems. In fact, for normalisation to be provable as is standardly done by an induction over the complexity of maximal formulas, the systems require modification. In a nutshell, the problem is that as they stand, once $\invertediota$ terms are added, applying the reduction procedures for maximal formulas with quantifiers and $\invertediota$ may produce maximal formulas of unbounded complexity because $\invertediota$ terms are substituted for free variables. To solve this problem I transpose an observation of Andrzej Indrzejczak's relating to sequent calculi for free logics without definite descriptions \citep[Sec. 4]{andrzejcutfreefreelogic} to natural deduction.\footnote{Indrzejczak notes the significance of his observation for the proof theory of definite descriptions in the conclusion to the paper cited. His cut elimination theorems for sequent calculi for various theories of definite descriptions are independent of it \citep{andrzejfregean, andrzejfreedefdescr, andrzejrussellian}.} The natural deduction version of a rule of Indrzejczak's permits the restriction of the rules for the quantifiers and $\invertediota$ to free variables or constants in such a way that only these are replaced for free variables in the reduction procedures. This ensures that the complexity of any new maximal formulas is less than that of the maximal formula removed, and induction can proceed as usual. 

I shan't consider a suitable subformula property for the systems with $\invertediota$. As Tennant's rules introduce and eliminate the $\invertediota$ operator in terms flanking $=$, this would require too many exceptions. 

Although the present paper focuses on formal issues, there will be occasion to comment briefly on the philosophical upshot of the results. Normalisation theorems are of interest in themselves, but they also have a wider philosophical significance. In proof-theoretic semantics, the theory that the meanings of logical expressions are defined by the rules of inference governing them, it is regarded as a necessary condition for them to do so that they permit normalisation.\footnote{This approach originates ultimately with a fertile remark of Gentzen's \citep[\S 5.13]{gentzenuntersuchungen}. It was the basis for Prawitz's \emph{inversion principle} and his proofs of normalisation for classical and intuitionist logic. The view has been developed in detail by Dummett as part of a theory of meaning (e.g. \citep[Chs 11-13]{dummettLBM}), a project to which Prawitz contributed, too (e.g. \citep{prawitzmeaningviaproofs}). The term `proof-theoretic semantics' was coined by Schroeder-Heister, who also made important contributions to the field. See \citep{schroederheistersemantics} for an overview.} 

Given the significance of Russell's theory of definite description, it is a little surprising that there is next to no discussion of the theory of definite descriptions in proof-theoretic semantics. Proof-theoretic semantics so far has rarely considered free logic and largely been restricted to sentence-forming operators and quantifiers. Term-forming operators hardly feature in the literature at all. A notable exception is Neil Tennant's work.\footnote{See \citep{tennantnatural} and \citep{tennantabstraction}. Tennant prefers the intuitionist version of the systems discussed in this paper. Strictly speaking, he would prefer an intuitionist core-logical version \citep{tennantcorelogic}, but this book does not discuss definite descriptions. For Tennant's take on meaning and proof theory, see also \citep{tennanteternal}.} The present paper is a contribution to the further development of a proof-theoretic semantics for definite descriptions and term-forming operators in general.

\section{Systems of Intuitionist and Classical Free Logic}
\subsection{Preliminaries}
The definition of the language is standard, except that there is a special one-place predicate $\exists!$ and in some systems the operator $\invertediota$. As usual in proof-theoretic investigation, I assume that there is an unlimited stock of variables, called \emph{parameters}, used as free variables, distinct in shape from those whose occurrences are bound by quantifiers: $a, b \ldots$ for the former, $x, y, z\ldots$ for the latter. Constants are designated by $c, d\ldots$, predicate letters by $P, Q, R\ldots$, identity by $=$, and the `existence predicate', indicating that a term refers, by $\exists !$. $A, B\ldots F, G \ldots$ range over formulas. If necessary, subscripts are used. The connectives are $\bot$, $\land$, $\lor$, $\rightarrow$, $\exists$ and $\forall$. $\neg A$ is defined as $A\rightarrow \bot$. $\invertediota$ takes a variable and a formula $F$ and forms a singular term out of them, $\invertediota xF$, where $x$ is bound by $\invertediota x$ in $F$. The terms of the language are the constants, parameters and the $\invertediota$ terms, the former two \emph{atomic}, the latter \emph{complex}. $t$ ranges over terms. $A_t^x$ names the formula that results from replacing the term $t$ for all free occurrences of $x$ in $A$. The distinction between parameters and bound variables ensures that replacements of terms in formulas is always possible. 

\begin{definition}
\normalfont\emph{Prime formulas} are those formed from parameters and constants by predicate letters, $\exists!$ or =. \emph{Atomic formulas} are those formed from terms by these expressions.  
\end{definition}

\noindent Deductions are defined as usual as certain kinds of trees labeled with formulas. I follow Troelstra's and Schwichtenberg's conventions for the discharge or closing of assumptions \citep[Sec 2.1.1]{troelstraschwichtenberg}: assumptions are assigned assumption classes, rules close or discharge all formulas in an assumption class, indicated by square brackets around the formulas in that class and a numeral to their right and to the right of the inference at which these assumptions are closed or discharged. Empty assumption classes, for vacuous discharge, are permitted. I'll use $\Pi$, $\Xi$, $\Sigma$ to stand for deductions, often, as is customary, displaying their conclusions below and some assumptions on top. 

\begin{definition}
\normalfont $\Gamma\vdash_S A$ means that there is a deduction of $A$ in formal system $S$ from some of the formulas in $\Gamma$ as open assumptions. 
\end{definition}

\subsection{Intuitionist Negative Free Logic}
These are the rules of Tennant's system of intuitionist negative free logic $\mathbf{INF}$ \citep[Sec. 7.10]{tennantnatural}:

\begin{center}
\AxiomC{$A$} 
\AxiomC{$B$}
\LeftLabel{$(\land  I)$ \ }
\BinaryInfC{$A\land  B$}
\DisplayProof\qquad\qquad 
\AxiomC{$A\land  B$}
\LeftLabel{$(\land  E)$ \ }
\UnaryInfC{$A$}
\DisplayProof\qquad 
\AxiomC{$A\land  B$} 
\UnaryInfC{$B$}
\DisplayProof
\end{center}

\begin{center}
\AxiomC{$[A]^i$}
\noLine
\UnaryInfC{$\Pi$}
\noLine
\UnaryInfC{$B$}
\RightLabel{$_i$}
\LeftLabel{$(\rightarrow I)$ \ }
\UnaryInfC{$A\rightarrow B$}
\DisplayProof \qquad \qquad
\AxiomC{$A\rightarrow B$}
\AxiomC{$A$}
\LeftLabel{$(\rightarrow E)$ \ }
\BinaryInfC{$B$}
\DisplayProof
\end{center}

\begin{center}
\AxiomC{$A$}
\LeftLabel{$(\lor I)$ \ }
\UnaryInfC{$A\lor B$}
\DisplayProof\qquad
\AxiomC{$B$}
\UnaryInfC{$A\lor B$}
\DisplayProof\qquad\qquad
\AxiomC{$A\lor B$}
\AxiomC{$[A]^i$}
\noLine
\UnaryInfC{$\Pi$}
\noLine
\UnaryInfC{$C$}
\AxiomC{$[B]^{j}$}
\noLine
\UnaryInfC{$\Sigma$}
\noLine
\UnaryInfC{$C$}
\RightLabel{$_{i, j}$}
\LeftLabel{$(\lor E)$ \ }
\TrinaryInfC{$C$}
\DisplayProof
\end{center}

\begin{prooftree}
\AxiomC{$\bot$}
\LeftLabel{$(\bot E)$ \ }
\UnaryInfC{$B$}
\end{prooftree}

\begin{center}
\AxiomC{$[\exists !a]^i$}
\noLine
\UnaryInfC{$\Pi$}
\noLine
\UnaryInfC{$A_a^x$}
\LeftLabel{$(\forall I)$ \ }
\RightLabel{$_i$}
\UnaryInfC{$\forall x A$}
\DisplayProof\qquad\qquad
\AxiomC{$\forall xA$}
\AxiomC{$\exists !t$}
\LeftLabel{$(\forall E)$ \ }
\BinaryInfC{$A_t^x$}
\DisplayProof
\end{center} 

\noindent where in $(\forall I)$, $a$ does not occur in $\forall xA$ nor in any undischarged assumptions of $\Pi$ except those in the assumption class of $\exists !a$. 

\begin{center}
\AxiomC{$A_t^x$}
\AxiomC{$\exists !t$}
\LeftLabel{$(\exists I)$ \ }
\BinaryInfC{$\exists x A$}
\DisplayProof\qquad\qquad
\AxiomC{$\exists xA$}
\AxiomC{$[A_a^x]^i \ [\exists !a]^j$}
\noLine
\UnaryInfC{$\Pi$}
\noLine
\UnaryInfC{$C$}
\RightLabel{$_{i, j}$}
\LeftLabel{$(\exists E)$ \ }
\BinaryInfC{$C$}
\DisplayProof
\end{center} 

\noindent where in $(\exists E)$, $a$ is not free in $C$, nor in $\exists xA$, nor in any undischarged assumptions of $\Pi$ except those in the assumption classes of $A_a^x$ and $\exists ! a$. 

\begin{center}
\AxiomC{$\exists ! t$}
\LeftLabel{$(= I^n)$ \ }
\UnaryInfC{$t=t$}
\DisplayProof\qquad\qquad
\AxiomC{$t_1=t_2$}
\AxiomC{$A_{t_1}^x$}
\LeftLabel{$(= E)$ \ } 
\BinaryInfC{$A_{t_2}^x$}
\DisplayProof
\end{center} 

\begin{prooftree}
\AxiomC{$Rt_1\ldots t_n$}
\LeftLabel{$(AD)$ \ }
\UnaryInfC{$\exists ! t_i$}
\end{prooftree}

\noindent where $R$ is an $n$-place predicate letter (but not $\exists!$) or identity and $1\leq i\leq n$. 

The superscript $n$ of $(=I^n)$ indicates that this is in the introduction rule for identity in negative free logic. $(AD)$, the rule of \emph{atomic denotation}, is typical for negative free logic: an atomic sentence can only be assertible or true if all terms that occur in it refer.  

I will often omit labels of rules in deductions, especially when they are simple and the rules mentioned in the surrounding text, but I'll add them where I think this helps understanding, especially when deductions are more complex and the rules involved less familiar.  

$\exists !$ is in one sense redundant, as what is often called \emph{Hintikka's Law} holds:\footnote{Hintikka equates existence, or rather his interpretation of Quine's dictum that to be is to be the value of a bound variable, with the right hand side of the displayed formula \citep[132f]{hintikkaexential}.}\bigskip

$\vdash \exists ! t\leftrightarrow \exists x \ x=t$\bigskip

\noindent In another sense it makes good sense to keep it primitive, if the meanings of the quantifiers are to be defined by the rules governing them, as they are by Tennant: without it, the definition of the meaning of the existential quantifier would be circular.

\subsection{Intuitionist Negative Free Logic with a Definite Description Operator}\label{iotarules}
The system $\mathbf{INF}^{\invertediota}$ results by adding the $\invertediota$ operator and Tennant's introduction and elimination rules for it to $\mathbf{INF}$ \citep[Sec. 7.10]{tennantnatural}: 

\begin{prooftree} 
\AxiomC{$\exists ! t$}
\AxiomC{$[a=t]^i$}
\noLine
\UnaryInfC{$\Xi$}
\noLine
\UnaryInfC{$F_a^x$}
\AxiomC{$[F_a^x]^j \ [\exists !a]^k$}
\noLine
\UnaryInfC{$\Pi$}
\noLine
\UnaryInfC{$a=t$}
\LeftLabel{$(\invertediota I)$ \ }
\RightLabel{$_{i, j, k}$}
\TrinaryInfC{$\invertediota xF=t$}
\end{prooftree} 

\noindent where $a$ does not occur in $\invertediota xF$, nor in $t$, nor in any undischarged assumptions except those in the assumption classes of $a=t$ in $\Xi$ or of $F_a^x$ and $\exists ! a$ in $\Pi$.\bigskip 

\begin{center} 
\AxiomC{$\invertediota xF=t$}
\AxiomC{$u=t$}
\LeftLabel{$(\invertediota E_1)$ \ } 
\BinaryInfC{$F_u^x$}
\DisplayProof\quad
\AxiomC{$\invertediota xF=t$}
\AxiomC{$F_u^x$}
\AxiomC{$\exists ! u$}
\LeftLabel{$(\invertediota E_2)$ \ }
\TrinaryInfC{$u=t$}
\DisplayProof\quad
\AxiomC{$\invertediota xF=t$}
\LeftLabel{$(\invertediota E_3)$ \ } 
\UnaryInfC{$\exists !t$}
\DisplayProof
\end{center} 

\noindent \label{ADandiotaE3}$(\invertediota E_3)$ is a special case of the rule $(AD)$. But it is properly regarded as an elimination rule for $\invertediota$, for, as we'll see, there is a reduction procedure for maximal formulas of the form $\invertediota xF=t$ that have been concluded by $(\invertediota I)$ and are premises of $(\invertediota E_3)$. So whenever $(AD)$ could be applied to the conclusion of $(\invertediota I)$, that is, when it is concluded that the right term refers, this is regarded as an application of $(\invertediota E_3)$. If it is the left term, the rule is $(AD)$. 

The rules for $\invertediota$ are what is now often called \emph{Lambert's Axiom} in rule form: 

\lbp{LA}{$LA$}{$\forall z (\invertediota xAx=z\leftrightarrow \forall y(Ay\leftrightarrow z=y))$}

\noindent $\mathbf{INF}^{\invertediota}$ is thus an intuitionist version of the minimal theory of definite descriptions in negative free logic.\footnote{Lambert developed his theory in \citep{lambertnotesEII} and \citep{lambertnotesEIII}, which are incorporated and expanded in \citep{lambertEThe}. Hintikka put forward a theory of definite descriptions around the same time \citep{hintikkatowardsdefdesc}. The label `minimal theory' is from \citep{lambertfoundations}. } 

A second axiom of Lambert's, sometimes called `cancellation', 

\lbp{CA}{$CA$}{$\invertediota x (x=t)=t$}

\noindent is provable conditionally on the existence of $t$ by an application of $(\invertediota I)$ in which $F$ is $x=t$ and $\exists !a$ in the rightmost subdeduction is discharged vacuously: 

\begin{prooftree} 
\AxiomC{$\exists ! t$}
\AxiomC{$[a=t]^1$}
\AxiomC{$[a=t]^1$}
\RightLabel{$_1$}
\TrinaryInfC{$\invertediota x (x=t)=t$}
\end{prooftree}

\subsection{Classical Negative Free Logic with and without a Definite Description Operator}
$\mathbf{CNF}$ has the introduction and elimination rules for $\rightarrow$, $\forall$ and $=$, $(AD)$ and $(\bot E_C)$:  

\begin{prooftree}
\AxiomC{$[\neg A]^i$}
\noLine
\UnaryInfC{$\Pi$}
\noLine
\UnaryInfC{$\bot$}
\RightLabel{$_i$}
\LeftLabel{$(\bot E_C) \ $}
\UnaryInfC{$A$}
\end{prooftree}

\noindent Vacuous discharged being permitted, $(\bot E)$ is a special case of $(\bot E_C)$. I'll refer to these applications of $(\bot E_C)$ by the former label. 

$\mathbf{CNF}^{\invertediota}$ results by adding $\invertediota$ and its rules to $\mathbf{CNF}$.

\subsection{Two Simplifying Lemmas}
The two lemmas in this section absolve us from having to consider certain tiresome cases in the normalisation theorem. 

\begin{lemma}\label{identityrestriction}
$(=E)$ may be restricted to atomic conclusions. 
\end{lemma}

\begin{proof} 
Well known and standard: break up a non-atomic formula by applying elimination rules until atomic formulas are reached, apply $(=E)$, reconstitute it by applying introduction rules.
\end{proof}

\noindent Lemma \ref{identityrestriction} holds for all the systems to be considered. More interesting and as far as I know new to the literature is the following: 

\begin{lemma}\label{botrestriction}
$(\bot E)$ may be restricted to prime conclusions. 
\end{lemma}

\begin{proof} 
(i) Evidently it is unnecessary to conclude $\bot$ from $\bot$ by $(\bot E)$. (ii) By a well known and standard result, $(\bot E)$ may be restricted to atomic conclusions. Thus it suffices to consider the case where the conclusion of $(\bot E)$ is an atomic formula. We'll first reduce these cases to identities flanked by only one $\invertediota$ term and then treat those. 

(iii) Let  $G$ be either $\exists !$ or $=$ flanked by two $\invertediota$ terms or an $n+m$-place predicate letter, $1<n, 0\leq m$, forming a formula with $n$ $\invertediota$ terms and $m$ atomic terms, the latter left implicit below: 

\begin{prooftree}
\AxiomC{$\bot$}
\UnaryInfC{$G(\invertediota x_1F_1\ldots \invertediota x_nF_n)$}
\end{prooftree}

\noindent These cases can all be reduced to applications of $(\bot E)$ to identities flanked by only one $\invertediota$ term by the following method. Take $n$ fresh parameters $a_1\ldots a_n$ and infer $a_1=\invertediota x_1F_1 \ldots a_n=\invertediota x_nF_n$ and $G(a_1\ldots a_n)$ by $n+1$ applications of $(\bot E)$, then apply $(=E)$ $n$ times to deduce $G(\invertediota x_1F_1\ldots \invertediota x_nF_n)$: 

\begin{prooftree}
\AxiomC{$\bot$}
\UnaryInfC{$a_n=\invertediota x_nF_n$}
\AxiomC{$\bot$}
\UnaryInfC{$a_2=\invertediota x_2F_2$}
\AxiomC{$\bot$}
\UnaryInfC{$a_1=\invertediota x_1F_1$}
\AxiomC{$\bot$}
\UnaryInfC{$G(a_1\ldots a_n)$}
\BinaryInfC{$G(\invertediota x_1F_1, a_2 \ldots a_n)$}
\BinaryInfC{$G(\invertediota x_1F_1, \invertediota x_2F_2\ldots a_n)$}
\noLine
\UnaryInfC{$\vdots$}
\noLine
\UnaryInfC{$G(\invertediota x_1F_1, \invertediota x_2F_2\ldots \invertediota x_{n-1}F_{n-1}, a_n)$}
\BinaryInfC{$G(\invertediota x_1F_1, \invertediota x_2F_2\ldots \invertediota x_{n-1}F_{n-1}, \invertediota x_nF_n)$}
\end{prooftree}

\noindent When $G$ is an identity flanked by two $\invertediota$ terms, $\invertediota xF=\invertediota yH$, there is a shorter method, as $a=\invertediota xF, a=\invertediota yH\vdash \invertediota xF=\invertediota yH$, but the method above works, too. 

(iv) This leaves the case of an identity flanked by one $\invertediota$ term. As $=$ is symmetric, it suffices to consider the case: 

\begin{prooftree}
\AxiomC{$\bot$}
\UnaryInfC{$\invertediota xF=b$}
\end{prooftree}

\noindent where $b$ is a an atomic term. Apply $(\invertediota I)$ with vacuous discharge, where $a$ is a fresh parameter: 

\begin{prooftree}
\AxiomC{$\bot$}
\UnaryInfC{$\exists !b$}
\AxiomC{$\bot$}
\UnaryInfC{$F_a^x$}
\AxiomC{$\bot$}
\UnaryInfC{$a=b$}
\TrinaryInfC{$\invertediota xF=b$}
\end{prooftree}

\noindent $F$ may be a complex formula or $\bot$ and contain further $\invertediota$ terms, so to establish the lemma, parts (i), (ii), (iii) and (iv) may need to be applied again. (i) is trivial: delete an application of $(\bot E)$ that concludes $\bot$ immediately whenever it arises as part of the procedure. The method for establishing (ii) reduces the number of connectives and quantifiers in conclusions of $(\bot E)$. The methods of (iii) and (iv) reduce in addition the number of $\invertediota$s in conclusions of $(\bot E)$. The result, therefore, can be established by an induction over the complexity of conclusions of $(\bot E)$. For the purposes of this lemma, let the degree for a formula be the sum of connectives, quantifiers and $\invertediota$s in it. The measure $\langle d, e\rangle$, ordered lexicographically, will do, where $d$ is the highest degree of any conclusion of $(\bot E)$, and $e$ is the number of conclusions of $(\bot E)$ of highest degree. Applying any of the procedures in (ii), (iii) or (iv) either reduces $e$ or, if there is only one conclusion of $(\bot E)$ of highest degree, reduces $d$. 
\end{proof}

\noindent Lemma \ref{botrestriction} holds for $\mathbf{INF}$, $\mathbf{INF}^{\invertediota}$ and carries over to $\mathbf{CNF}$ and $(\bot E_C)$: its conclusion, too, can be restricted to prime formulas (here the same as the atomic ones). 

Lemma \ref{botrestriction} does not carry over to $(\bot E_C)$ in $\mathbf{CNF}^{\invertediota}$: when discharge is not vacuous, conclusions of $(\bot E_C)$ cannot be restricted to prime, but only to atomic conclusions: conclusions containing $\invertediota$ terms must be admitted, but only atomic ones. It would go too far to give a rigorous proof of this result here, and in any case, a further restriction is not needed for the normalisation proof to be given later. The following should suffice to elucidate the reason why the result fails. To show that $(\bot E_C)$ need not be applied to conclude a complex formula $A$, the usual procedure is to apply an elimination rule for the main operator of $A$, derive $\bot$ by assuming the negation of a subformula of $A$, conclude $\neg A$, apply $(\bot E_C)$ to the subformula, and to do so as often as required to apply an introduction rule for the main operator of $A$ to conclude $A$ and discharge any auxiliary assumptions. This method does not work here. Suppose $\Gamma, \neg \ \invertediota xF=a\vdash\bot$. We could take a fresh parameter $b$ and conclude $F_b^x$ by $(\invertediota E_1)$ from $\invertediota xF=a$ and $b=a$, assume $\neg F_b^x$ to derive $\bot$, and so deduce $\neg \ \invertediota xF=a$; thus $\Gamma, b=a\vdash F_b^x$. We could apply $(\invertediota E_2)$ to $\invertediota xF=a, F_b^x$ and $\exists !b$ to derive $b=a$, assume $\neg \ b=a$ to derive $\bot$ and so deduce $\neg \ \invertediota xF=a$; thus $\Gamma, F_b^x, \exists !b\vdash b=a$. Thus we have the two left deductions required for an application of $(\invertediota I)$. But we do not have $\exists !b$, and there is no way to derive it from the material we are given. Besides, even with $\exists !b$ we'd only be able to derive $\invertediota xF=b$, not $\invertediota xF=a$, as $(\invertediota I)$ requires a fresh parameter. The situation is no better for atomic formulas other than identities. To avoid concluding $P(\invertediota xF)$ we'd need a formula $\invertediota xF=a$ to conclude $\neg P(\invertediota xF)$ from $\neg P(a)$, and there is no way to discharge it. 

Henceforth I shall assume all applications of $(=E)$, $(\bot E)$ and $(\bot E_C)$ to be restricted according to Lemmas \ref{identityrestriction} and \ref{botrestriction}. That is, in $\mathbf{CNF}^{\invertediota}$, applications of $(\bot E_C)$ with vacuous discharge are assumed to have prime conclusions, just as for applications of $(\bot E)$ in $\mathbf{INF}^{\invertediota}$.

\section{Preliminaries to Normalisation}
\subsection{General Notions}\label{general}
As usual, I assume that in any application of rules with restrictions on parameters the parameter is introduced into the deduction solely for the purpose of that application of the rule, occurring nowhere else.\footnote{Troelstra and Schwichtenberg call this the \emph{pure variable condition} \citep[Sec. 2.1.2]{troelstraschwichtenberg}. The terminology goes back to Prawitz \citeyearpar[28f]{prawitznaturaldeduction}. Gentzen used the neat term \emph{Eigenvariable} \citep[186]{gentzenuntersuchungen}: each application of such a rule has its own variable.} This ensures that in the transformations applied to deductions in the process of normalisation, no `clashes' of variables can occur. $\Pi_t^a$ names the deduction that results from replacing the term $t$ for all occurrences of the parameter $a$ in $\Pi$. \AxiomC{$\Pi$}
\noLine
\UnaryInfC{$[A]$}
\noLine
\UnaryInfC{$\Sigma$}
\DisplayProof indicates that $\Pi$ is used to conclude all assumptions in the assumption class of $A$ in $\Sigma$. This notation is used to indicate that formulas discharged by a rule in one deduction are instead concluded  by another rule in a transformed deduction. 

\begin{definition}\normalfont
The \emph{major premise} of an elimination rule is the formula that displays the connective or $\invertediota$ in the general statement of the rule, here always the leftmost premise. All others are \emph{minor premises}. 
\end{definition}

\noindent Applications of $(\lor E)$ and $(\exists E)$ give rise to sequences of formulas of the same shape, all minor premises and conclusions of $(\lor E)$ or $(\exists E)$, except the first and the last ones: the first is only a minor premise, the last only a conclusion. It is convenient to capture this situation in a way that covers every formula in a deduction (cf. \citep[49]{prawitznaturaldeduction} and \citep[179]{troelstraschwichtenberg}: 

\begin{definition}\label{segment}\normalfont
A \emph{segment} is a sequence of formulas $C_1\ldots C_n$ such that $C_1$ is not the conclusion of $(\lor E)$ or $(\exists E)$, $C_n$ is not a minor premise of $(\lor E)$ or $(\exists E)$, and if $n>1$ then for all $i<n$, $C_i$ is a minor premise of $(\lor E)$ or $(\exists E)$, $C_{i+1}$ the conclusion. $n$ is the \emph{length} of the segment. 
\end{definition}

\noindent I'll say that $C$ is on a segment and speak of segments as being  premises, conclusions, discharged assumptions of rules depending on whether their last or first formulas are.  

A segment of length $1$ is a formula, and I shall often refer to them as such. Occasionally I shall use the pleonasm `formula or segment'.  

Definition \ref{segment} ignores a minor issue. Call an application of $(=E)$ \emph{vacuous} if the major premise is $t=t$. The definition ignores the possibility that the formulas on a segment are minor premises and conclusions of vacuous applications of $(=E)$. However, as then the minor premise and the conclusion of $(=E)$ are identical, they can evidently be removed from deductions without loss and without trouble. Including this possibility is a needless complication. Vacuous applications of $(=E)$ can arise as a result of the transformations of deductions in normalisation procedures. I shall assume that they are always removed together with the procedure. I shall leave this largely implicit, but as a reminder that they may occur, I shall at various points mention vacuous applications of $(=E)$. 

\begin{definition}\label{maxformseg}
\normalfont A segment is \emph{maximal} if its last formula is the major premise of an elimination rule, and if its length is $1$, it is the conclusion of an introduction rule. 
\end{definition}

\noindent Accordingly a \emph{maximal formula} is a formula occurrence that is the conclusion of an introduction rule and the major premise of an elimination rule. 

The degree of a formula is normally defined as the number of logical symbols occurring in it. For our purposes, however, quantifiers need to count for two because of the existence assumptions in their rules. This ensures that $\forall xA$ and $\exists xA$ always have a higher complexity than the premises and discharged assumptions of their introduction rules. This is not needed for $\invertediota$, as counting $\invertediota$ and $=$ as one each, $\invertediota xF=t$ always has a higher complexity than any premise or discharged assumption of $(\invertediota I)$ (at least 2). The degree $d(A)$ of a formula $A$ and $d(t)$ of terms is defined by simultaneous induction as follows: 

\begin{definition}\label{degree}
\normalfont
(a) If $A$ is a prime formula $Rt_1\ldots t_n$, then $d(A)=0$ if $R$ is a predicate letter, and $d(A)=1$ if $R$ is $\exists !$ or $=$. (b) $d(t)=0$, if $t$ is an atomic term, $d(F)+1$ if $t$ is $\invertediota xF$. (c) If $A$ is an atomic formula $Rt_1\ldots t_n$, then $d(A)=d(t_1)+\ldots d(t_n)$ if $R$ is a predicate letter, and $d(A)=d(t_1)+\ldots d(t_n)+1$ if $R$ is $\exists!$ or $=$. (d) If $A$ is: 

(i) $\bot$, then $d(A)=1$; 

(ii) $\neg B$, then $d(A)=d(B)+1$; 

(iii) $B\land C$, $B\supset C$ or $B\lor C$, then $d(A)=d(B)+d(C)+1$; 

(iv) $\forall xB$ or $\exists xB$, then $d(A)=d(B)+2$
\end{definition}

\noindent The degree of a segment is the degree of the formula on it.

\subsection{Failure of the Subformula Property}
\noindent As I won't consider the subformula property for systems with $\invertediota$, the definition of \emph{subformula} is standard:

\begin{definition}\label{subformula}
\normalfont $A$ is a subformula of $A$; $A$ is a subformula of $\neg A$; $A, B$ are subformulas of $A\land B$, $A \lor B$, $A \rightarrow B$; and for any atomic term $t$, $A_t^x$ is a subformula of $\exists xA$, $\forall xA$; and if $A$ is a subformula of $B$ and $B$ is a subformula of $C$, then $A$ is a subformula of $C$. 
\end{definition}

\noindent The subformula property is usually defined as follows: 

\begin{definition}\label{subformulaproperty}
\normalfont A deduction has the \emph{subformula property} if every formula occurring on it is a subformula of the conclusion or of some open assumption. 
\end{definition}

\noindent This fails already in standard first-order logic with identity. For instance: 

\begin{prooftree}
\AxiomC{$t_3=t_4$}
\AxiomC{$t_1=t_2$}
\AxiomC{$Rt_1t_3$}
\LeftLabel{$_{(=E)}$}
\BinaryInfC{$Rt_2t_3$}
\LeftLabel{$_{(=E)}$}
\BinaryInfC{$Rt_1t_4$}
\end{prooftree}

\noindent $Rt_2t_3$ is not a subformula of any other formula, and there is no way to rearrange the deduction to change this. It also fails due to rules specific to negative free logic. For instance, to deduce $A_t^x$ from $\forall xA$ requires $\exists !t$, which we may be able to infer from an atomic formula $Pt$: 

\begin{prooftree}
\AxiomC{$\forall xA$}
\AxiomC{$Pt$}
\UnaryInfC{$\exists !t$}
\BinaryInfC{$A_t^x$}
\end{prooftree}

\noindent $\exists !t$ need not be a subformula of another formula, and there may be no way to rearrange this deduction. Identity gives rise to further cases, e.g. to conclude $t=t$ by $(=I^n)$ requires $\exists !t$, which may be concluded from $Pt$. 

This circumscribes where exceptions to the subformula property are found. I will later define a restricted version that holds for deductions in $\mathbf{INF}$ and $\mathbf{CNF}$.\bigskip 

\noindent \textsc{Comment}. I am not regarding $(AD)$ as an introduction rule for $\exists!$, nor $(\exists I)$, $(\forall E)$ and $(=I^n)$ as elimination rules for it. There are philosophical questions that arise here, which I will discuss briefly on p.\pageref{jaskowski}f. and p.\pageref{referenceandexistence}. An extended discussion must wait for another occasion.\footnote{Some thoughts on this issue are in \citep{kurbisiotaIII}.}

\subsection{Maximal Segments Specific to Negative Free Logic}
Maximal formulas arise because of detours in deductions, to use Gentzen's phrase. Normalisation theorems show that these detours can be avoided. They are unnecessary to derive the conclusion. The process of normalisation may also remove unnecessary assumptions from the deduction. The thought is that a proof in normal form appeals only to what is essential to prove the conclusion. 

The rules of negative free logic give rise to detours in addition to those of definition \ref{maxformseg}. For instance, $(=I^n)$ followed by $(AD)$ and sometimes conversely:   

\begin{center}
\AxiomC{$\exists !t$}
\LeftLabel{$_{(=I^n)}$}
\UnaryInfC{$t=t$}
\LeftLabel{$_{(AD)}$}
\UnaryInfC{$\exists !t$}
\DisplayProof\qquad
\AxiomC{$t=t$}
\LeftLabel{$_{(AD)}$}
\UnaryInfC{$\exists !t$}
\LeftLabel{$_{(=I^n)}$}
\UnaryInfC{$t=t$}
\DisplayProof
\end{center}

\noindent This is clearly unnecessary. Similarly when $t=t$ and $\exists!t$ are stretched out by $(\lor E)$ or $(\exists E)$. In the systems with $\invertediota$, $(\invertediota E_3)$ could be applied instead of $(AD)$, but this clearly makes no difference, as we can just rename the rule. 

\begin{definition}\label{maxexistsidentitysegment}
\normalfont A \emph{maximal $=$-segment} is a segment (of formulas $t=t$) such that its first formula is concluded by $(=I^n)$ and its last is the premise of $(AD)$. A \emph{maximal $\exists!$-segment} is a segment (of formulas $\exists !t$) such that its first formula is concluded from $t=t$ by $(AD)$ and its last is the premise of $(=I^n)$. 
\end{definition}

\noindent In $\mathbf{CNF}$ and $\mathbf{CNF}^{\invertediota}$ we may speak of maximal $\exists!$- and $=$-formulas. 

Recall that we are ignoring vacuous applications of $(=E)$. This absolves us from considering the case that a maximal $=$-segment contains what might be called \emph{totally vacuous} applications of $(=E)$, those where all three formulas of the inference are $t=t$. 

$(=E)$ gives rise to sequences of formulas quite similar to segments, except that terms are replaced in the formulas constituting the segment: 

\begin{definition}
\normalfont An $(=E)$-\emph{segment} is a sequence of segments $\sigma_1\ldots \sigma_n$ such that $\sigma_1$ is the minor premise but not the conclusion of $(=E)$, $\sigma_n$ is the conclusion but not the minor premise of $(=E)$, and if $1<i<n$, $\sigma_i$ is the conclusion of $(=E)$ and the minor premise of $(=E)$. 
\end{definition}

\noindent I'll refer to the major premises to the left of formulas on an $(=E)$-segment as the major premises of the segment. 

$(=E)$-segments can give rise to unnecessary detours, if $(AD)$ is applied to their last formulas or if their formulas have the form $\exists!t$. For example: 

\begin{center}
\begin{flushleft}(a) $i$ is $3$ or $4$:\end{flushleft}\AxiomC{$t_2=t_4$}
\AxiomC{$t_1=t_2$}
\AxiomC{$t_1=t_3$}
\BinaryInfC{$t_2=t_3$}
\BinaryInfC{$t_4=t_3$}
\UnaryInfC{$\exists ! t_i$}
\DisplayProof

\begin{flushleft}(b) $i$ is $4, 5$ or $6$:\end{flushleft}\AxiomC{$t_3=t_6$}
\AxiomC{$t_2=t_5$}
\AxiomC{$t_1=t_4$}
\AxiomC{$Rt_1t_2t_3$}
\BinaryInfC{$Rt_4t_2t_3$}
\BinaryInfC{$Rt_4t_5t_3$}
\BinaryInfC{$Rt_4t_5t_6$}
\UnaryInfC{$\exists ! t_i$}
\DisplayProof

\begin{flushleft}(c)\end{flushleft}\AxiomC{$t_3=t_4$}
\AxiomC{$t_2=t_3$}
\AxiomC{$t_1=t_2$}
\AxiomC{$\exists !t_1$}
\BinaryInfC{$\exists !t_2$}
\BinaryInfC{$\exists !t_3$}
\BinaryInfC{$\exists !t_4$}
\DisplayProof
\end{center}

\noindent These constructions involve terms that may not be needed to derive the conclusion. In the systems without $\invertediota$, these would be atomic terms that name objects the existence of which may be irrelevant to proving the conclusion (by $(AD)$, for any $t_j$ in (a) and (b), $\exists !t_j$). Free logic being concerned with the avoidance of existence assumptions, this is undesirable. In the systems with $\invertediota$, there may in addition be complex terms that refer to objects by predicates that not are needed to prove the conclusion: this would be a case of introducing unnecessary concepts into the proof. 

The $(=E)$-segments can be omitted:\bigskip

\noindent (a) If $i=3$, $\exists !t_3$ could have been concluded by $(AD)$ from the minor premise of the first application of $(=E)$; if $i=4$, $\exists !t_4$ could have been concluded by $(AD)$ from the major premise of the last application of $(=E)$.\bigskip

\noindent (b) $\exists !t_i$ could have been concluded by $(AD)$ from one of the major premises of $(=E)$; if one or more of $t_4, t_5$ or $t_6$ happens to be the same as $t_1, t_2$ or $t_3$, $\exists !t_i$ could also have been concluded from the first formula of the segment.\bigskip

\noindent (c) $\exists !t_4$ could have been concluded by $(AD)$ from the major premise of the last application of $(=E)$.\bigskip

\noindent Clearly this situation generalises, and also to cases with $(\lor E)$ or $(\exists E)$ interspersed. As before, in the systems with $\invertediota$, $(\invertediota E_3)$ could be applied instead of $(AD)$, but this again makes no difference, and we just rename the rule. 

\begin{definition}\label{=Esegment}
\normalfont An $(=E)$-segment is \emph{maximal} if its last segment is premise of $(AD)$ or if all formulas on it are formed from $\exists !$ and a term. 
\end{definition}

\noindent To distinguish the maximal segments defined in this section from those defined in the last section, I'll refer to the latter by `maximal $I/E$-segments' (for introduction/elimination). By `maximal segment' I usually mean both, but sometimes only $I/E$-segments, if it is clear from context what is meant. 

The reduction procedures that remove the new maximal segments from deductions are as follows:\bigskip

\noindent I. For maximal $\exists!$- and $=$-segments: remove the application of the rule that concludes the segment, the segment and the formula concluded from it (i.e. we proceed directly from the formula from which the segment is concluded to the rule applied to the formula concluded from the segment).\bigskip

\noindent II. For maximal $(=E)$-segments: (i) If the formula on the segment is formed from $\exists!$ and a term, conclude its last formula from the major premise of the last application of $(=E)$, removing all the rest. (ii) If not, conclude $\exists !t_i$ from either the first formula of the segment or from the lowest major premise of an application of $(=E)$ that contains $t_i$, removing all the rest. 

\begin{lemma}\label{removenewsegments}
Any deduction can be transformed into one without maximal $\exists!$-, $=$- and $(=E)$-segments. 
\end{lemma}

\begin{proof} Procedures I and II reduce the number of applications of rules in the deduction. None is added. Applying them will therefore come to an end. We can proceed in the following way. (1) Recall that vacuous applications of $(=E)$ are always removed. (2) Remove maximal $=$- and $\exists!$-segments. Carrying out procedure I does not introduce new maximal $=$-, $\exists!$- or $(=E)$-segments. The latter is evident. Removing a maximal $=$-segment could only introduce a maximal $\exists!$-segment if the segment from which it is concluded and the segment concluded from it are concluded by $(AD)$ and premise of $(=I^n)$ respectively, in which case they are both already maximal $\exists !$-segments. Then two maximal $\exists!$-segments are fused into one (with length one less than their sum). Similarly when removing maximal $\exists!$-segments. The procedure shortens any such sequence of maximal $=$- and $\exists!$-segments, but as they have the same formulas first and last, the entire sequence can also be removed at once. (3) Three cases are to be considered when removing maximal $(=E)$-segments. (a) Both parts of procedure II can introduce a new maximal $=$-segment if the last or lowest major premise of $(=E)$ is $t_i=t_i$ or if the first formula of the maximal $(=E)$-segment has this form. If the former, the application of $(=E)$ is vacuous, and the problem is avoided by removing vacuous applications of $(=E)$. If the latter, the first formula is the conclusion of $(=I^n)$ and the last formula of the maximal $(=E)$-segment is the premise of $(AD)$, so the problem is solved by removing everything between the premise of $(=I^n)$ and the conclusion of $(AD)$ (i.e. move straight from the premise of $(=I^n)$ to the rule applied to the conclusion of $(AD)$). (b) Both parts of procedure II can introduce a new maximal $\exists!$-segment if the last formula of the $(=E)$-segment is the premise of $(=I^n)$. In the case of part (i), the major premise of the last application of $(=E)$ would then have to have the form $t=t$, in which case the application is vacuous and the problem solved by removing it. In the case of part (ii), either the first formula of the maximal $(=E)$-segment or the lowest major premise is $t_i=t_i$. If the latter, the application of $(=E)$ is vacuous, so remove it. If the former, $\exists ! t_i$ is concluded by $(AD)$ from premise $t_i=t_i$, so the problem is solved by removing everything between the premise of $(AD)$ and the conclusion of $(=I^n)$ (i.e. move straight from the premise of $(AD)$ to the rule applied to the conclusion of $(=I^n)$). (c) It can create a maximal $(=E)$-segment if one of the major premises is the last formula of an $(=E)$-segment. Then continue the process there. Consider the entire cluster of $(=E)$-segments, that is, the maximal $(=E)$-segment, all $(=E)$-segments concluding any of its major premises, and all $(=E)$-segments concluding any of their major premises, etc. By removing the maximal $(=E)$-segment the number of applications of $(=E)$ has been reduced in this cluster, hence the process comes to an end. 
\end{proof}

\subsection{Normal Form and Rank of Deductions}
The definition of normal form is standard: 

\begin{definition}\label{normalform}
\normalfont A deduction is in \emph{normal form} if it contains no maximal segments. 
\end{definition}

\noindent Normalisation is proved by induction over the complexity of deductions, where maximal $\exists!$-, $=$- and $(=E)$-segments are not counted, as they are taken care of by Lemma \ref{removenewsegments}: 

\begin{definition}\label{rank}
\normalfont The \emph{rank} of a deduction is the pair $\langle d, l\rangle$, where $d$ is the highest degree of a maximal I/E-segment or $0$ if there is none, and $l$ is the sum of the lengths of I/E maximal segments of highest degree. $\langle d, l\rangle < \langle d', l'\rangle$ iff either (i) $d<d'$ or (ii) $d=d'$ and $l<l'$. 
\end{definition}

\section{Normalisation and Subformula Property for $\mathbf{INF}$}
\subsection{Normalisation}
The reduction procedures for removing maximal segments of length 1 (i.e. formulas) with the sentential connectives as main operators are standard and won't be repeated: they are those given by Prawitz \citeyearpar[35ff]{prawitznaturaldeduction}. The permutative reduction  procedures for shortening maximal segments of length longer than 1 are standard, too: for $(\lor E)$ they are those given by Prawitz \citeyearpar[51]{prawitznaturaldeduction}, for $(\exists E)$ a trivial variation of those given by him. The reduction procedures that remove applications of $(\lor E)$ and $(\exists E)$ in which vacuous discharge occurs are also as usual: applications of $(\lor E)$ in which no or only one assumption is discharged are evidently superfluous, similarly if no assumption is discharged by $(\exists E)$. Applications of that rule that discharge only one assumption, however, are not superfluous and not removed, unless the major premise is derived by $(\exists I)$. 

The reduction procedures for maximal formulas with quantifiers as main operators do not pose much of a problem either, but Tennant omits them from \citep{tennantnatural}, and Troelstra and Schwichtenberg do not consider normalisation of a corresponding system \citep[6.5]{troelstraschwichtenberg}, so here they are. Replace the inferences on the left by those on the right: 

\begin{center}
\AxiomC{$[\exists !a]^i$}
\noLine
\UnaryInfC{$\Pi$}
\noLine
\UnaryInfC{$A_a^x$}
\UnaryInfC{$\forall xA$}
\AxiomC{$\Sigma$}
\noLine
\UnaryInfC{$\exists !t$}
\RightLabel{$_i$}
\BinaryInfC{$A_t^x$}
\DisplayProof\qquad$\leadsto$\qquad
\AxiomC{$\Sigma$}
\noLine
\UnaryInfC{$[\exists !t]$}
\noLine
\UnaryInfC{$\Pi_t^a$}
\noLine
\UnaryInfC{$A_t^x$}
\DisplayProof

\bigskip

\AxiomC{$\Xi$}
\noLine
\UnaryInfC{$A_t^x$}
\AxiomC{$\Sigma$}
\noLine
\UnaryInfC{$\exists !t$}
\BinaryInfC{$\exists x A$}
\AxiomC{$[A_a^x]^i\ [\exists !a]^j$}
\noLine
\UnaryInfC{$\Pi$}
\noLine
\UnaryInfC{$C$}
\RightLabel{$_{i, j}$}
\BinaryInfC{$C$}
\DisplayProof\qquad$\leadsto$\qquad
\def\defaultHypSeparation{\hskip .01in}
\AxiomC{$\Xi$}
\noLine
\UnaryInfC{$[A_t^x]$}
\AxiomC{$\Sigma$}
\noLine
\UnaryInfC{$[\exists !t]$}
\noLine
\BinaryInfC{$\Pi_t^a$}
\noLine
\UnaryInfC{$C$}
\DisplayProof
\end{center}
 
\begin{theorem}
Deductions in $\mathbf{INF}$ can be brought into normal form. 
\end{theorem}

\begin{proof}
By an induction over the rank of deductions by applying the reduction procedures for maximal segments. The methods of Prawitz \citeyearpar[50]{prawitznaturaldeduction} and Troelstra and Schwichtenberg \citeyearpar[182]{troelstraschwichtenberg} work here, too. Prawitz chooses a maximal segment of highest degree such that no maximal segment of highest degree stands above it or above a minor premise to its right or has an element that is such a minor premise. Troelstra and Schwichtenberg choose the rightmost maximal segment of highest degree that has no maximal segment of highest degree standing above it. We check that the reduction procedures lower the rank of deductions. But first apply Lemma \ref{removenewsegments}. Hence the reduction procedures cannot increase the lengths of maximal $(=E)$-segments, as they were all removed. For the sentential connectives, the situation is as for intuitionist logic, except that maximal $=$-, $\exists !$- or $(=E)$-segments may be created. So apply Lemma \ref{removenewsegments} afterwards.  The following cases need to be considered regarding the reduction procedures for the quantifiers. (a) They could introduce a maximal formula $A_t^x$ or, if it is already on a maximal segment, increase its length. In both cases the rank of the deduction is lowered because the degree of $A_t^x$ is lower than that of the maximal formula removed. (b) (i) If $\exists !t$ is the conclusion of $(AD)$ in $\Sigma$ and $\exists !a$ is premise of $(=I^n)$ in $\Pi$ the procedure introduces a maximal $\exists$-segment. (ii) If $\exists !t$ is conclusion of $(=E)$ in $\Sigma$ and $\exists !a$ is premise of $(AD)$ in $\Pi$ the procedure introduces a maximal $(=E)$-segment. Both cases are dealt with by applying Lemma \ref{removenewsegments} immediately after the reduction step. 
\end{proof}

\subsection{Subformula Property}
The following modifies Prawitz's notion of a path slightly to fit $\mathbf{INF}$: 

\begin{definition}\label{path}
\normalfont A \emph{path} is a sequence of formulas $A_1\ldots A_n$ such that 

\noindent (a) $A_1$ is an assumption not discharged by $(\lor E)$ of $(\exists E)$; 

\noindent (b) if $A_i$ is any premise of an introduction rule other than $(\exists I)$, the left premise of $(\exists I)$, the premise of $(AD)$ or $(=I^n)$, the minor premise of $(=E)$ or the major premise of an elimination rule other than $(\lor E)$, $(\exists E)$ or $(=E)$, then $A_{i+1}$ is the conclusion of the rule; 

\noindent (c) if $A_i$ is the major premise of $(\lor E)$ or $(\exists E)$, then $A_{i+1}$ is an assumption discharged by the rule; 

\noindent (d) $A_n$ is either the minor premise of $(\rightarrow E)$ or $(\forall E)$, the right premise of $(\exists I)$, the major premise of $(=E)$ or the conclusion of the deduction.
\end{definition}

\noindent Paths are naturally divided into segments. Indeed, in the definition above, `segment' could replace `formula'. I will speak of paths being in deductions and of formulas and segments being on paths. 

\begin{definition}\normalfont
A path that ends in the conclusion of a deduction has order $0$; a path has order $n+1$ if its last formula ends to the left or right of a formula on a path with order $n$. 
\end{definition}

\begin{corollary}\label{majorpreceding}
On a path in a deduction in $\mathbf{INF}$ in normal form major premises of elimination rules precede conclusions of introduction rules. 
\end{corollary}

\begin{proof}
Suppose there is a conclusion $A$ of an introduction rule on a path. Let $\sigma_1$ be the segment beginning with $A$. If the path does not end with $\sigma_1$, then (1) $\sigma_1$ cannot be the major premise of $(=E)$, nor the right premise of $(\forall E)$ or $(\exists I)$; (2) as $A$ is the conclusion of an introduction rule, $\sigma_1$ cannot be the minor premise of $(=E)$, nor the premise of $(=I^n)$, $(AD)$ or $(\bot E)$; (3) as the deduction is normal, it cannot be the major premise of an elimination rule. Hence it can only be premise of an introduction rule for a sentential connective or the left premise of $(\forall I)$ or $(\exists I)$. But the same applies to any other segment, if the path continues. Hence for any path on which there is a conclusion of an introduction rule, if there are also major premises of elimination rules on it, they must precede the conclusions of introduction rules. 
\end{proof}

\begin{corollary}\label{formofpaths}
A path in a deduction in $\mathbf{INF}$ in normal form begins with a (possibly empty) sequence of major premises of elimination rules (only formulas, not segments), which is followed either by the last segment of the path or by a sequence of premises of $(AD)$ or $(=I^n)$ or minor premises of $(=E)$ or by a premise of an introduction rule, and ends in a (possibly empty) sequence of conclusions of introduction rules. 
\end{corollary} 

\begin{proof}
By Corollary \ref{majorpreceding}, any major premises of elimination rules precede the introduction rules. The premises of $(=I^n)$, $(=E)$ and $(AD)$ are atomic and so cannot come before major premises of elimination rules. For the same reason, they cannot come after any introduction rules. This leaves the place in between.
\end{proof}

\noindent The first part of a path may be called the \emph{E-part}, the last the \emph{I-part}, the one in the middle the \emph{M-part}. Notice that the M-part is never empty. The following corollary establishes a result about the form of M-parts of paths in deductions in normal form. It is here that $(=E)$ and the rules specific to negative free logic, $(=I^n)$ and $(AD)$, are applied, and because of normal form, these applications are only very limited and follow a certain order. For instance, if all three rules just mentioned are applied, they must be applied in the order $(AD)$, $(=I^n)$, $(=E)$, where ($=I^n)$ concludes the minor premise of $(=E)$, and there follow no more applications of $(AD)$ or $(=I^n)$. 

\begin{corollary}\label{M-part}
(i) On the M-part of a path in a deduction in normal form: 

(a) There is at most one application of $(AD)$; 

(b) There is at most one application of $(=I^n)$; 

(c) If there is an application of $(AD)$ and an application of $(=I^n)$, $(AD)$ precedes $(=I^n)$ and there are no applications of $(=E)$ between them. 

\medskip

(ii) The first segment of an M-part of a path in a deduction in normal form may be: 

(a) the first segment of an $(=E)$-segment: then its last segment is the last segment of the M-part; 

(b) the premise of $(AD)$: then its conclusion is either the last segment of the M-part, or it is the premise of $(=I^n)$, in which case the premise of $(AD)$ is different from the conclusion of $(=I^n)$, and the conclusion of $(=I^n)$ is either the last segment of the M-part or first premise of an $(=E)$-segment the last conclusion of which is the first segment of the M-part; 

(c) the premise of $(=I^n)$: then its conclusion is either the last segment of the M-part or the first premise of an $(=E)$-segment the last conclusion of which is the last segment of the M-part; 

(d) the last segment of the M-part.
\end{corollary}

\begin{proof}
(i) (a) The conclusion of $(AD)$ has the form $\exists !t$. To apply it again we need a formula that is either an identity or formed from a predicate letter. The only rules that could be applied are $(\lor E)$, $(\exists E)$, $(=E)$ and $(=I^n)$. The first two conclude $\exists !t$ again and $(=E)$ concludes a formula of the same form and there'd be a maximal $(=E)$-segment, which is excluded as the deduction is in normal form. Applying $(=I^n)$ permits us to conclude a formula to which $(AD)$ may be applied, but doing so would create a maximal $\exists !$-segment, which is excluded on paths of deductions in normal form. (b) follows for similar reasons: to apply $(=I^n)$ twice, the second application requires a premise of the form $\exists !t$, and this could only be brought along by an application of $(AD)$, which would create a maximal $=$-segment. (c) follows because if $(=I^n)$ preceded $(AD)$ there would be a maximal $=$-segment, and if there were applications of $(=E)$ in between $(AD)$ and $(=I^n)$, there would be a maximal  $(=E)$-segment. 

(ii) (a) If there was an application of $(AD)$ later, this would be a maximal $(=E)$-segment, contradicting normality. If there was an application of $(=I^n)$, the formulas on the $(=E)$-segment would have the form $\exists ! t$, also contradicting normality. (b) The conclusion of $(AD)$ is of form $\exists !t$ and hence by normality no $(=E)$ can follow immediately. If the premise of $(AD)$ was the same as the conclusion of $(=I^n)$, we'd have a maximal $\exists!$-segment. If a $(=E)$-segment follows after $(=I^n$), it follows from (a) that it ends the path. (c) follows from clause (i). (d) requires no argument. 
\end{proof}

\begin{corollary}\label{subformulas}
(i) Any formula in the E-part of a path in a deduction in normal form is a subformula of the immediately preceding formula. 

(ii) Any formula in the I-part of such a path is a subformula of its immediate successor. 

(iii) The first formula of the M-part is either the first formula of the path or a subformula of the last formula of the E-part; the last formula of the M-part is either the last formula of the path or a subformula of the first formula of the I-part. Formulas in between may not be subformulas of any other formulas. 
\end{corollary}

\begin{proof}
(i), (ii) and (iii) are evident by inspection of the rules and Corollary \ref{M-part}. 
\end{proof}

\begin{definition}\normalfont
A deduction has the \emph{free subformula property} if all formulas are subformulas of the undischarged assumptions or of the conclusion of the deduction, except possibly formulas of the form $t_1=t_2$ or $Rt_1\ldots t_n$ on $(=E)$-segments; formulas of the form $\exists!t$ that are minor premises of $(\forall E)$ or right premises of $(\exists I)$; or formulas of the form $\exists !a$ that are discharged by $(\forall I)$ or $(\exists E)$. 
\end{definition}

\begin{corollary}\label{freesubformula}
Deductions in $\mathbf{INF}$ in normal form have the free subformula property. 
\end{corollary}

\begin{proof}
By inspection of the rules and an induction over the order of paths. For paths of order $0$ this follows immediately from Corollary \ref{subformulas}. Evidently the exempt formulas all and only occur on M-part of paths. Suppose the corollary holds for paths of order $n$ and consider a path $\pi$ of order $n+1$. There are four ways in which a path can end. 

(I) $\pi$ ends in a minor premise of $(\rightarrow E)$. There are two options. 

(A) $\pi$ has an E-part. Then apart from any formulas between the first and the last of the M-part, which are exempt, the situation is as in intuitionist logic. The last formula of $\pi$ is a subformula of a formula of a path of lower order, hence any formulas on $\pi$'s I-part and the last formula of the M-part are subformulas of a path of lower order. The first formula of the M-part is a subformula of the last formula of $\pi$'s E-part, and all formulas on the E-part are either subformulas of an undischarged assumption of the deduction or, if they are subformulas of a discharged assumption of the deduction, they are discharged by $(\rightarrow E)$, and hence are subformulas of a formula on a path of lower order. (As there is an E-part, they cannot be discharged by $(\forall I)$. 

(B) $\pi$ has no E-part. Then the first formula of the M-part is either an undischarged assumption of the deduction, or it is discharged by $(\rightarrow E)$, or it has the form $\exists !a$ and is discharged by $(\forall I)$ or $(\exists E)$ (it can't be $A_t^x$, as then there'd be an E-part). In the first two cases, we're done. In the other two cases, $\exists !a$ is exempt, but we need to consider any other formulas on the path. $\exists !a$ can only be premise of $(=I^n)$ or the last formula of the M-part, hence only options (ii) (c) and (d) of Corollary \ref{M-part} can be the case. If (ii) (d), there are four options. $\exists !a$ is on a segment that is (1) minor premise of $(\forall I)$, (2) right premise of $(\exists E)$, (3) minor premise of $(\rightarrow E)$ or (4) premise of an introduction rule. (1) and (2) are exempt. If (3), it is a subformula of a formula on a path of lower order. If (4), $\pi$ ends in a minor premise of $(\rightarrow E)$, and again it is a subformula of a formula on a path of lower order. 

If (ii) (c), there are two options. (1) If there is no $(=E)$-segment, the conclusion of $(=I^n)$ is on the last segment of the M-part, which can only be minor premise of $(\rightarrow E)$ or premise of an introduction rule, and hence it is a subformula of a formula on a path of lower order. (2) If there is an $(=E)$-segment, the formulas on it are exempt, and its last formula is either on a segment that is minor premise of $(\rightarrow E)$ or premise of an introduction rule, and the situation is as before. 

(II) $\pi$ ends in the major premise of $(=E)$. Then $\pi$ has no I-part and the last formula of its M-part is $t_1=t_2$. There are two options. (1) $t_1=t_2$ is also the first formula of $\pi$'s M-part. If it is an undischarged assumption of the deduction or discharged by $(\rightarrow E)$, we're done. If it is discharged by $(\lor E)$ or $(\exists E)$, it is a subformula of a formula on the E-part of $\pi$, and hence a subformula of the first formula on $\pi$, and thus either of an undischarged assumption or one discharged by $(\rightarrow I)$ and again we're done. (2) $t_1=t_2$ is not the first formula of $\pi$'s M-part. As the deduction is normal, it cannot have been concluded by $(=I^n)$, and hence can only be the last formula of an $(=E)$-segment or concluded by an elimination rule. If the latter, we're done as in previous cases. If the former, it and all formula on the $(=E)$-segment are exempt, but the usual reasoning applies to the first formula of the $(=E)$-segment. 

(III) Here we consider the two options that $\pi$ ends in a minor premise of $(\forall I)$ or the right premise of $(\exists E)$. Then $\pi$ has no I-part, and there are two options for its M-part. (1) It consists only of a segment $\exists !t$. (2) $\exists !t$ is concluded by $(AD)$. (The option that it is concluded by $(=E)$ is excluded by normality). (1) If $\exists !t$ is an undischarged assumption, discharged by $(\rightarrow I)$, or if $\pi$ has an E-part, we're done, for reasons as before; if it is discharged by ($\forall I)$ or $(\exists E)$ (right discharged assumption), it is exempt. (2) Then it is exempt, and reasoning as before applies to the premise of $(AD)$. 
\end{proof}

\noindent In relation to the usual definition of subformula property we have:  

\begin{corollary}
Any exceptions to the subformula property in deductions in normal form in $\mathbf{INF}$ occur between the first and the last formula of the M-part of paths. 
\end{corollary}

\begin{proof}
Immediate from Definition \ref{subformulaproperty} and Corollary \ref{subformulas}. 
\end{proof}

\section{A System inspired by Ja\'skowski}
It would be possible to loosen the restriction on the subformula property for deductions in normal form a little. The proof of Corollary \ref{freesubformula} shows that formulas $(\exists !a)$ that are premises of $(\forall E)$ or $(\exists I)$ are often subformulas of the undischarged premises or conclusions of deductions in normal form, namely if they are concluded by elimination rules or discharged by $(\lor E)$ or $(\rightarrow I)$ or if they take the place of the discharged assumption $A_a^x$ of $(\exists E)$. Similarly for those discharged by $(\forall I)$ or $(\exists E)$, if they are premises of introduction rules. We could also count formulas of the form $\exists !t$ that are the minor premise of $(\forall E)$ as subformulas of its major premise and those that are the right premise of $(\exists I)$ as subformulas of its conclusion, in analogy with $(\rightarrow E)$ and $(\land I)$; and analogously counting assumptions $\exists !a$ discharged by $(\forall I)$ and $(\exists E)$ as subformulas of the conclusion and the major premise, respectively. This makes some sense, as after all the quantifiers are supposed to carry existential import and range only over what exists. But this still leaves occurrences of formulas of the form $\exists !t$ that are not subformulas of any formulas on the deduction, namely those concluded by $(AD)$ and the premises of $(=I^n)$ and variations thereof.  

A more elegant option with a better result goes back to Ja\'skowski \citep[Sec. 5]{jaskowskirules}. Eschew use of $\exists !$ altogether, permit terms to occur in rules, and reformulate the rules in which it occurred accordingly:

\begin{center}
\AxiomC{$[a]^i$}
\noLine
\UnaryInfC{$\Pi$}
\noLine
\UnaryInfC{$A_a^x$}
\LeftLabel{$(\forall I^J)$ \ }
\RightLabel{$_i$}
\UnaryInfC{$\forall x A$}
\DisplayProof\qquad\qquad
\AxiomC{$\forall xA$}
\AxiomC{$t$}
\LeftLabel{$(\forall E^J)$ \ }
\BinaryInfC{$A_t^x$}
\DisplayProof
\end{center} 

\noindent where in $(\forall I)$, $a$ does not occur in $\forall xA$ nor in any undischarged assumptions of $\Pi$ except those in the assumption class of $a$. 

\begin{center}
\AxiomC{$A_t^x$}
\AxiomC{$t$}
\LeftLabel{$(\exists I^J)$ \ }
\BinaryInfC{$\exists x A$}
\DisplayProof\qquad\qquad
\AxiomC{$\exists xA$}
\AxiomC{$[A_a^x]^i \ [a]^j$}
\noLine
\UnaryInfC{$\Pi$}
\noLine
\UnaryInfC{$C$}
\RightLabel{$_{i, j}$}
\LeftLabel{$(\exists E^J)$ \ }
\BinaryInfC{$C$}
\DisplayProof
\end{center} 

\noindent where in $(\exists E)$, $a$ is not free in $C$, nor in $\exists xA$, nor in any undischarged assumption of $\Pi$ except those in the assumption classes of $A_a^x$ and $a$. 

\begin{center}
\AxiomC{$t$}
\LeftLabel{$(= I^{nJ})$ \ }
\UnaryInfC{$t=t$}
\DisplayProof\qquad\qquad
\AxiomC{$Rt_1\ldots t_n$}
\LeftLabel{$(AD^J)$ \ }
\UnaryInfC{$t_i$}
\DisplayProof
\end{center}

\noindent where $R$ is an $n$-place predicate letter or identity and $1\leq i\leq n$.\bigskip

\noindent Call the reformulated system $\mathbf{INF}^J$. 

$\mathbf{INF}^J$ loses none of the expressiveness of $\mathbf{INF}$: as observed, $\vdash \exists !t\leftrightarrow \exists x \ x=t$, so $\exists!$ is in this sense redundant. $(=E)$-segments in which the minor premises and conclusions have the form $\exists !t$ are no longer possible, but the normalisation theorem showed them to be superfluous.

Terms appearing as premises or conclusions of the $J$-rules also occur in their conclusions or other premises. This gives a notion of a subformula property: 

\begin{definition}
\normalfont A deduction has the \emph{free term subformula property} if all formulas and terms on it are or occur in subformulas of its undischarged assumptions or its conclusion, except possibly formulas of the form $t_1=t_2$ or $Rt_1\ldots t_n$ on $(=E)$-segments.
\end{definition}

\noindent The normalisation proof goes through exactly as before, except that it is a little simpler, as one kind of maximal $(=E)$-segments need no longer be considered: 

\begin{theorem}
Deductions in $\mathbf{INF}^J$ can be brought into normal form. 
\end{theorem}

\noindent The form of paths in deductions in normal form stays \emph{mutatis mutandis} the same as established in corollaries \ref{majorpreceding}, \ref{formofpaths}, \ref{M-part} and \ref{subformulas}, and so: 

\begin{corollary}
Deductions in normal form in $\mathbf{INF}^J$ have the free term subformula property.
\end{corollary}

\noindent For the classical version $\mathbf{CNF}^J$ similar results hold as to be established for $\mathbf{CNF}$. 

It should be possible to go even further by adopting an approach to identity formalised by Indrzejczak, where the role of terms is assimilated entirely to that of formulas, and identity statements are proved on the basis of rules for terms. Then the subformula property holds even in the presence of identity. See \citep{indrzejczakequality}. Pursuing this further must be left for another occasion.\bigskip

\noindent \textsc{Comment.}\label{jaskowski} If the rules for the quantifiers are to define their meanings, a thesis of proof-theoretic semantics, an explanation of the use of terms as premises and conclusions in the $J$-rules is required. Adopting Ja\'skowski's account lends itself to a neat approach to addressing this issue, and it has an interesting historical antecedent as a philosophical foundation. Ja\'skowski introduces a symbol analogous to the sign of supposition of propositions to be placed in front of variables (although he preferred to call them `arbitrary constants'). This marks that the referent of the variable, not otherwise defined, is kept constant throughout the reasoning to follow. It corresponds to the phrase `Consider an arbitrary $x$' used in proofs. Like the domains of suppositions of propositions, the `domain of constancy', as Ja\'skowski calls it, of a variable can be closed by an application of a rule, in his case the introduction rule for the universal quantifier. Ja\'skowski has no primitive rules for the existential quantifier, but evidently they are exactly analogous to those for the universal quantifier. A philosophical foundation for this approach can be found in Brentano. Textor develops an explanation of the meaning of the existence predicate on the basis of Brentano's account of acknowledging or positing objects \citep{textorneobrentanianexistence}.\footnote{\citep{kurbistextorcomment} is a comment on a presentation of Textor's paper before publication.} This is a non-propositional attitude: thinkers acknowledge or posit objects in thought. Acknowledging or positing \emph{that} something exists is to be explained as a propositional attitude derivative thereof. Textor motivates the use of an existence predicate in natural deduction as part of his account, and this could be used to motivate the rules of $\mathbf{INF}$. But the attitude of acknowledging or positing objects also lends itself to be incorporated directly into the rules to motivate those of $\mathbf{INF}^J$. Using $t$ as a premise in $(\forall E^J)$ and $(\exists I^J)$ indicates that $t$ has been acknowledged: only terms referring to acknowledged objects are legitimate terms to use in these rules. An object can only be asserted to be self-identical if it has been acknowledged, which is $(=I^{nJ})$. Some rules permit the discharge of such acknowledgments: their conclusions hold no matter which object has been acknowledged, as is the case in $(\forall I^J)$ and $(\exists E^J)$: they no longer commit to the acknowledgement. `Positing an object' is a neat description of what happens if an assumption is made for the sake of discharge by $(\forall I)$ and $(\exists E)$. Finally, some propositions commit one to acknowledging objects, e.g. atomic propositions, which motivates $(AD^J)$.

\section{Adding Definite Descriptions}
\subsection{A Problem Solved by a New Rule for Identity}
For a normalisation theorem to be provable by induction over the complexity of formulas, we must count $\invertediota$ terms in addition to connectives and quantifiers, as was already done in Definition \ref{degree}. This creates a problem with the reduction procedures for maximal formulas of form $\forall x A$ and $\exists xA$: if $t$ is a complex term, they may introduce maximal formulas of higher degree than those removed. There is no apparent systematic way of avoiding this, e.g. by applying the reduction procedures to a suitably chosen formula. Replacing a parameter by an $\invertediota$ term increases the complexity of the formula. Looking only at a maximal formula of highest degree, we do not know whether the replacement of a parameter by an $\invertediota$ term will not turn a maximal formula that had a lower than maximal degree before into one the degree of which is now higher. 

The problem is most straightforwardly solved by an observation made by Indrzejczak \citeyearpar[Sec. 4]{andrzejcutfreefreelogic} in relation to a number of free logics formulated in cut free sequent calculi: $(\forall E)$ and $(\exists I)$ can be restricted to atomic instantiating terms, given the rule of the next lemma, which is derivable in $\mathbf{INF}$. 

\begin{lemma}\label{=genintroderivable}
This rule is derivable given $(=I^n$), $(\exists I)$ and $(\exists E)$: 
\end{lemma}

\begin{prooftree}
\AxiomC{$\exists !t$}
\AxiomC{$[a=t]^i$}
\noLine
\UnaryInfC{$\Pi$}
\noLine
\UnaryInfC{$C$}
\LeftLabel{$(=I^{nG})$}
\RightLabel{$_i$}
\BinaryInfC{$C$}
\end{prooftree}

\noindent where the parameter $a$ does not occur in $t$, $C$ nor any open assumptions of $\Pi$ other than those of the assumption class of $a=t$.

\begin{proof}
As observed earlier, $\exists !t\vdash \exists x \ x=t$, apply $(\exists E)$.
\end{proof}

\noindent $(=I^{nG})$ is a new introduction rule for $=$. It has the form of what Negri and von Plato \citeyearpar[217]{negriplatostructural} and Milne \citeyearpar{milneinversion,milnesubformula} call \emph{general introduction rules}. In its presence $(=I^n)$ is redundant: 

\begin{lemma}\label{=introderivable}
Given $(=E)$ and $(=I^{nG})$, $(=I^n)$ is derivable. 
\end{lemma}

\begin{proof}\phantom{r}\vspace{-0.4cm}
\begin{prooftree}
\AxiomC{$\exists !t$}
\AxiomC{$[a=t]^1$}
\LeftLabel{$_{(=E)}$}
\AxiomC{$[a=t]^1$}
\BinaryInfC{$t=t$}
\LeftLabel{$_{(=I^{nG})}$}
\RightLabel{$_1$}
\BinaryInfC{$t=t$}
\end{prooftree}
\end{proof}

\noindent The symmetry of identity will play a prominent role in the following, so we prove it here, which also gives another example of a use of $(=I^{nG})$: 

\begin{lemma}\label{symmetry}
$t_1=t_2\vdash t_2=t_1$.
\end{lemma}

\begin{proof}\phantom{r}\vspace{-0.4cm}
\begin{prooftree}
\AxiomC{$t_1=t_2$}
\LeftLabel{$_{(AD)}$}
\UnaryInfC{$\exists !t_1$}
\AxiomC{$t_1=t_2$}
\AxiomC{$[a=t_1]^1$}
\LeftLabel{$_{(=E)}$}
\BinaryInfC{$a=t_2$}
\AxiomC{$[a=t_1]^1$}
\LeftLabel{$_{(=E)}$}
\BinaryInfC{$t_2=t_1$}
\LeftLabel{$_{(=I^{nG})}$}
\RightLabel{$_1$}
\BinaryInfC{$t_2=t_1$}
\end{prooftree}
\end{proof}

\noindent Notice once more how the subformula property cannot be upheld: $\exists !t_1$, $a=t_1$ and $a=t_2$ are not subformulas of any undischarged premises or of the conclusion. 

Now for the point of introducing $(=I^{nG})$: 

\begin{lemma}\label{forallIexistsErestricted}
Given $(=I^{nG})$, $(\forall E)$ and $(\exists I)$ may be restricted to atomic terms. 
\end{lemma}

\begin{proof}
Any application of $(\exists I)$ and $(\forall E)$ where $t$ is complex can be replaced by the following, where $a$ is a fresh parameter: 

\begin{prooftree}
\AxiomC{$\exists !t$}
\AxiomC{$[a=t]^i$}
\AxiomC{$\forall xA$}
\AxiomC{$[a=t]^i$}
\LeftLabel{$_{(AD)}$}
\UnaryInfC{$\exists !a$}
\LeftLabel{$_{(\forall E)}$}
\BinaryInfC{$A_a^x$}
\LeftLabel{$_{(=E)}$}
\BinaryInfC{$A_t^x$}
\LeftLabel{$_{(=I^{nG})}$}
\RightLabel{$_i$}
\BinaryInfC{$A_t^x$}
\end{prooftree}

\begin{prooftree}
\AxiomC{$\exists !t$}
\AxiomC{$[a=t]^i$}
\AxiomC{$A_t^x$}
\LeftLabel{$_{(=E)}$}
\BinaryInfC{$A_a^x$}
\AxiomC{$[a=t]^i$}
\LeftLabel{$_{(AD)}$}
\UnaryInfC{$\exists !a$}
\LeftLabel{$_{(\exists I)}$}
\BinaryInfC{$\exists x A$}
\RightLabel{$_i$}
\LeftLabel{$_{(=I^{nG})}$}
\BinaryInfC{$\exists xA$}
\end{prooftree}
\end{proof}

\noindent The proof in fact shows something stronger:  $(\forall E)$ and $(\exists I)$ can be restricted to parameters, as $t$ might as well be a constant. But this is not needed in the following.\bigskip

\noindent \textsc{Comment.} Kürbis \citeyearpar{kurbisgenrulesint,kurbisgenrulescl} counts formulas discharged by general introduction rules that are the major premises of their elimination rules as maximal. There would, indeed, be a reduction procedure for such formulas, where \emph{all} maximal formulas arising from an application of $(=I^{nG})$ are replaced together after the following pattern: 

\begin{center}
\AxiomC{$\exists !t$}
\AxiomC{$[a=t]^i$}
\AxiomC{$\Sigma$}
\noLine
\UnaryInfC{$A_a^x$}
\BinaryInfC{$A_t^x$}
\noLine
\UnaryInfC{$\Pi$}
\noLine
\UnaryInfC{$C$}
\RightLabel{$_i$}
\BinaryInfC{$C$}
\DisplayProof
\quad$\leadsto$\quad
\alwaysNoLine
\AxiomC{$\Sigma_t^a\ast$}
\UnaryInfC{$A_t^x$}
\UnaryInfC{$\Pi_t^a\ast$}
\UnaryInfC{$C$}
\DisplayProof
\end{center}

\noindent If there are non-maximal formulas in the assumption class of $a=t$, then $\Pi_t^a\ast$ and $\Sigma_t^a\ast$ are the results of removing the ensuing vacuous applications of $(=E)$. But this won't work if $t$ is an $\invertediota$ term. As before, the replacement of $a$ by $t$ in $\Pi$ and $\Sigma$ may then create maximal formulas of unknown degree. What is more, the use to which $(=I^{nG})$ is put in handing normalisation in the presence of $\invertediota$ prevents us from removing Kürbis-style maximal formulas from deductions: this use will generate maximal formulas of exactly that kind. 

\label{referenceandexistence}The following philosophical questions arise. Should we not demand that these formulas be removable from deductions? Furthermore, the deduction that derives $(=I^{nG})$ in the proof of Lemma \ref{=genintroderivable} contains a maximal formula. So $(=I^{nG})$ comes at the cost of hiding a detour.\footnote{I owe this objection to a referee.} This is not the place to address these questions exhaustively, so a sketch of an answer must suffice. The answer to both questions lies, I think, in the connection between reference and existence. If a definite description $\invertediota xF$ refers, i.e. if $\exists !\invertediota xF$, then $(=I^{nG})$ permits the introduction of an \emph{ad hoc} name for its referent. We can, that is, baptise the object, to use Kripke's expression. This allows us to refer to it without having to describe it as an $F$. The significance of this comes out in modal contexts. The name is rigid, the definite description need not be. The move ensures that we can talk about the same object in every possible world. By contrast, we cannot baptise what does not exist; here all we have is the definite description. These are issues orthogonal to those of how the meanings of logical expressions are defined by their rules. So even if $(=I^{nG})$ hid or gave rise to unremovable maximal formulas, this would not upset the aims of proof-theoretic semantics. Further discussion of the wider issues touched upon here must wait for another occasion.

\subsection{Reduction Procedures for $\invertediota$ and Restrictions of its Rules}
The reduction procedures for identities flanked by an $\invertediota$ term, the first two taken from \citep[169]{tennantnatural}, are the following:\bigskip 

\noindent 1. $(\invertediota I)$ followed by $(\invertediota E_1)$: 

\begin{center} 
\AxiomC{$\Sigma_1$}
\noLine
\UnaryInfC{$\exists ! t$}
\AxiomC{$[a=t]^i$}
\noLine
\UnaryInfC{$\Xi$}
\noLine
\UnaryInfC{$F_a^x$}
\AxiomC{$[F_a^x]^j \ [\exists !a]^k$}
\noLine
\UnaryInfC{$\Pi$}
\noLine
\UnaryInfC{$a=t$}
%\LeftLabel{$(\invertediota I)$ \ }
\RightLabel{$_{i, j, k}$}
\TrinaryInfC{$\invertediota xF=t$}
\AxiomC{$\Sigma_2$}
\noLine
\UnaryInfC{$u=t$}
\BinaryInfC{$F_u^x$}
\DisplayProof
\quad $\leadsto$ \quad
\alwaysNoLine
\AxiomC{$\Sigma_2$}
\UnaryInfC{$[u=t]$}
\UnaryInfC{$\Xi_u^a$}
\UnaryInfC{$F_u^x$}
\DisplayProof
\end{center}

\noindent 2. $(\invertediota I)$ followed by $(\invertediota E_2)$: 

\begin{center}
\AxiomC{$\Sigma_1$}
\noLine
\UnaryInfC{$\exists ! t$}
\AxiomC{$[a=t]^i$}
\noLine
\UnaryInfC{$\Xi$}
\noLine
\UnaryInfC{$F_a^x$}
\AxiomC{$[F_a^x]^j \ [\exists !a]^k$}
\noLine
\UnaryInfC{$\Pi$}
\noLine
\UnaryInfC{$a=t$}
%\LeftLabel{$(\invertediota I)$ \ }
\RightLabel{$_{i, j, k}$}
\TrinaryInfC{$\invertediota xF=t$}
\AxiomC{$\Sigma_2$}
\noLine
\UnaryInfC{$F_u^x$}
\AxiomC{$\Sigma_3$}
\noLine
\UnaryInfC{$\exists ! u$}
%\LeftLabel{$(\invertediota E_2)$ \ }
\TrinaryInfC{$u=t$}
\DisplayProof\quad$\leadsto$\quad
\alwaysNoLine
\AxiomC{$\mathbin{\stackon[6pt]{[F_u^x]}{\Sigma_2}} \ \mathbin{\stackon[6pt]{[\exists !u]}{\Sigma_3}}$}
\UnaryInfC{$\Pi_u^a$}
\UnaryInfC{$u=t$}
\DisplayProof
\end{center}

\noindent 3. $(\invertediota I)$ followed by $(\invertediota E_3)$: 

\begin{center}
\AxiomC{$\Sigma_1$}
\noLine
\UnaryInfC{$\exists ! t$}
\AxiomC{$[a=t]^i$}
\noLine
\UnaryInfC{$\Xi$}
\noLine
\UnaryInfC{$F_a^x$}
\AxiomC{$[F_a^x]^j \ [\exists !a]^k$}
\noLine
\UnaryInfC{$\Pi$}
\noLine
\UnaryInfC{$a=t$}
%\LeftLabel{$(\invertediota I)$ \ }
\RightLabel{$_{i, j, k}$}
\TrinaryInfC{$\invertediota xF=t$}
%\LeftLabel{$(\invertediota E_3)$ \ } 
\UnaryInfC{$\exists !t$}
\DisplayProof\quad$\leadsto$\quad
\AxiomC{$\Sigma_1$}
\noLine
\UnaryInfC{$\exists ! t$}
\DisplayProof
\end{center} 

\noindent The same problem as for the quantifiers arises. If $u$ is a complex term, $\Xi_u^a$ and $\Sigma_u^a$ may contain maximal formulas of higher degree than the formula $\invertediota x F=t$ that is removed by the first and second reduction procedures. A new problem is that if $u$ is a complex term, $F_u^x$ can turn into a maximal formula of unknown degree. $u=t$, $\exists !u$ and $\exists !t$ pose no problem, as, if maximal, they are handled by Lemma \ref{removenewsegments}. 

This problem is solved by the following lemma, which also builds on an observation of Indrzejczak's \citeyearpar{andrzejrussellian}. In the following deductions, double lines mark inferences by symmetry of identity (Lemma \ref{symmetry}). 

\begin{lemma}\label{iotarestricted}
Given $(=I^{nG})$ and $(=E)$, $t$ and $u$ in $(\invertediota I)$, $(\invertediota E_1)$ and $(\invertediota E_2)$ can be restricted to atomic terms.
\end{lemma}

\begin{proof}
(1) Replace an application of $(\invertediota I)$ where $t$ is a complex term $\invertediota yG$ by: 

{\scriptsize
\begin{prooftree}
\def\defaultHypSeparation{\hskip .05in}
\AxiomC{$\exists !\invertediota yG$}
\AxiomC{$[b=\invertediota yG]^l$}
\AxiomC{$[b=\invertediota yG]^l$}
\LeftLabel{$_{(AD)}$}
\UnaryInfC{$\exists !b$}
\AxiomC{[$a=b]^i$}
\AxiomC{$[b=\invertediota yG]^l$}
\LeftLabel{$_{(=E)}$}
\BinaryInfC{$[a=\invertediota yG]$}
\noLine
\UnaryInfC{$\Xi$}
\noLine
\UnaryInfC{$F_a^x$}
\AxiomC{$[F_a^x]^j \ [\exists !a]^k$}
\noLine
\UnaryInfC{$\Pi$}
\noLine
\UnaryInfC{$a=\invertediota yG$}
\doubleLine
\UnaryInfC{$\invertediota yG=a$}
\AxiomC{$[b=\invertediota yG]^l$}
\doubleLine
\UnaryInfC{$\invertediota yG=b$}
\LeftLabel{$_{(=E)}$}
\BinaryInfC{$a=b$}
\LeftLabel{$_{(\invertediota I)}$}
\RightLabel{$_{i, j, k}$}
\TrinaryInfC{$\invertediota xF=b$}
\LeftLabel{$_{(=E)}$}
\BinaryInfC{$\invertediota xF=\invertediota yG$}
\LeftLabel{$_{(=I^{nG})}$}
\RightLabel{$_l$}
\BinaryInfC{$\invertediota xF=\invertediota yG$}
\end{prooftree} 
}

\noindent where $a$ and $b$ are fresh parameters. 

(2) (a) Replace an application of $(\invertediota E_1)$ where $t$ is a term $\invertediota yG$ and $u$ atomic by: 

{\footnotesize
\begin{prooftree}
\AxiomC{$\invertediota xF=\invertediota yG$}
\LeftLabel{$_{(AD)}$}
\UnaryInfC{$\exists !\invertediota yG$}
\AxiomC{$\invertediota xF=\invertediota yG$}
\doubleLine
\UnaryInfC{$\invertediota yG=\invertediota xF$}
\AxiomC{$[a=\invertediota yG]^i$}
\doubleLine
\UnaryInfC{$\invertediota yG=a$}
\LeftLabel{$_{(=E)}$}
\BinaryInfC{$\invertediota xF=a$}
\AxiomC{$[a=\invertediota yG]^i$}
\doubleLine
\UnaryInfC{$\invertediota yG=a$}
\AxiomC{$u=\invertediota yG$}
\LeftLabel{$_{(=E)}$}
\BinaryInfC{$u=a$}
\LeftLabel{$_{(\invertediota E_1)}$}
\BinaryInfC{$F_u^x$}
\LeftLabel{$_{(=I^{nG})}$}
\RightLabel{$_i$}
\BinaryInfC{$F_u^x$}
\end{prooftree}
}

\noindent where $a$ is a fresh parameter. 

(b) Replace an application of $(\invertediota E_1)$ where $u$ is a term $\invertediota zH$ and $t$ atomic by:

{\footnotesize
\begin{prooftree}
\AxiomC{$\invertediota zH=t$}
\LeftLabel{$_{(AD)}$}
\UnaryInfC{$\exists !\invertediota zH$}
\AxiomC{$\invertediota zH=t$}
\LeftLabel{$_{(AD)}$}
\UnaryInfC{$\exists !t$}
\AxiomC{$[b=\invertediota zH]^i$}
\AxiomC{$\invertediota xF=t$}
\AxiomC{$[b=t]^j$}
\LeftLabel{$_{(\invertediota E_1)}$}
\BinaryInfC{$F_b^x$}
\LeftLabel{$_{(=E)}$}
\BinaryInfC{$F_{\invertediota zH}^x$}
\LeftLabel{$_{(=I^{nG})}$}
\RightLabel{$_j$}
\BinaryInfC{$F_{\invertediota zH}^x$}
\LeftLabel{$_{(=I^{nG})}$}
\RightLabel{$_i$}
\BinaryInfC{$F_{\invertediota zH}^x$}
\end{prooftree}
}

\noindent where $b$ is a fresh parameter.

(c) If both $t$ and $u$ are complex, either carry out procedure (a) and replace the inference by $(\invertediota E_1)$ by the construction in (b), or carry out procedure (b) and replace the inference by the construction in (a). 

If both $t$ and $u$ are complex, there are also two alternative methods that lead to less complex deductions. (i) Say $u$ is $\invertediota xH$. Then replace $u$ in $(\invertediota E_1)$ by a fresh parameter $b$, so that it concludes $F_b^x$, derive $F_{\invertediota zH}^x$ from $b=\invertediota zH$ by $(=E)$, and discharge the former by an application of $(=I^{nG})$, with its premise derived from $\invertediota zH=\invertediota yG$ by $(AD)$. Then apply construction (a). (ii) Say $t$ is $\invertediota yG$. Then replace $t$ in $(\invertediota E_1)$ by a fresh parameter $a$ and discharge the ensuing formulas $\invertediota xF=a$ and $u=a$ by applications of $(=I^{nG})$ with their premises derived by $(AD)$. Then apply construction (b). 

(3) (a) Replace an application of $(\invertediota E_2)$ where $t$ is a term $\invertediota yG$ and $u$ atomic by
{\footnotesize
\begin{prooftree}
\AxiomC{$\invertediota xF=\invertediota yG$}
\LeftLabel{$_{(AD)}$}
\UnaryInfC{$\exists !\invertediota yG$}
\AxiomC{$[a=\invertediota yG]^i$}
\AxiomC{$[a=\invertediota yG]^i$}
\doubleLine
\UnaryInfC{$\invertediota yG=a$}
\AxiomC{$\invertediota xF=\invertediota yG$}
\LeftLabel{$_{(=E)}$}
\BinaryInfC{$\invertediota xF=a$}
\AxiomC{$F_u^x$}
\AxiomC{$\exists ! u$}
\LeftLabel{$_{(\invertediota E_2)}$}
\TrinaryInfC{$u=a$}
\LeftLabel{$_{(=E)}$}
\BinaryInfC{$u=\invertediota yG$}
\RightLabel{$_i$}
\BinaryInfC{$u=\invertediota yG$}
\end{prooftree}
}
\noindent where $a$ is a fresh parameter. 

(b) Replace an application of $(\invertediota E_2)$ where $u$ is a term $\invertediota yG$ and $t$ atomic by
{\scriptsize
\begin{prooftree}
\def\defaultHypSeparation{\hskip .1in}
\AxiomC{$\exists !\invertediota yG$}
\AxiomC{$[b=\invertediota yG]^i$}
\AxiomC{$\invertediota xF=t$}
\AxiomC{$[b=\invertediota yG]^i$}
\AxiomC{$F_{\invertediota yG}^x$}
\LeftLabel{$_{(=E)}$}
\BinaryInfC{$F_b^x$}
\AxiomC{$[b=\invertediota yG]^i$}
\LeftLabel{$_{(AD)}$}
\UnaryInfC{$\exists !b$}
\LeftLabel{$_{(=E)}$}
\TrinaryInfC{$b=t$}
\LeftLabel{$_{(=E)}$}
\BinaryInfC{$\invertediota yG=t$}
\RightLabel{$_i$}
\LeftLabel{$_{(=I^{nG})}$}
\BinaryInfC{$\invertediota yG=t$}
\end{prooftree}
}

(c) If both $t$ and $u$ are complex, either carry out procedure (a) and replace the inference by $(\invertediota E_2)$ by the construction in (b), or carry out procedure (b) and replace the inference by the construction in (a). 

Here, too, when both $t$ and $u$ are complex, there are methods that lead to less complex deductions, but this is left to the reader. 
\end{proof}

\noindent \textsc{Comment.}\label{explicitness} The restriction on the rules for $\invertediota$ may have philosophical significance. Do\v sen requires of an analysis of the meaning of an expression that a sentence in which it occurs only once is paraphrased by an equivalent sentence in which it does not occur \citep[369, 371f]{dosenpunctuation}. The reason is that otherwise it is not clear that the expression or rather a sequence of occurrances of that expression has been analysed.  This requirement is reminiscent of one often imposed on rules of sequent calculi, especially those intended to define the meanings of connectives by their inference rules. Wansing calls rules in which the connective they govern occurs only once and only in the conclusion \emph{explicit} (\citep[127]{wansingsequentmodal}, \citep[10]{wansingidea}). Although $(AD)$ cannot be so restricted, it may be motivated independently by a general requirement behind negative free logic: the truth or assertibility of an atomic proposition requires that all the terms occurring in it refer. Another criterion Wansing imposes is \emph{separation}: no other connective than the one they govern is mentioned in the rules. The rules for $\invertediota$ do not satisfy separation, as identity occurs in them, but requiring separation is rather too stringent. It is a legitimate procedure to define one expression in terms of another. The meaning of the second expression then depends on the first. In set theory, e.g., operations on the numbers are defined after the numbers have been defined. What should be avoided is that such dependencies are circular.

\subsection{Normalisation}
Due to these problems and their solutions I will prove normalisation not for $\mathbf{INF}^{\invertediota}$, but for the system $\mathbf{INF}^{\invertediota}{'}$ that results from it by replacing $(=I^n)$ by $(=I^{nG})$ and restricting $t$ in $(\forall E)$, $(\exists I)$ and $(\invertediota I)$, and $t$ and $u$ in $(\invertediota E_1)$ and $(\invertediota E_2)$ to atomic terms. These systems are equivalent: 

\begin{theorem}\label{equivalencerestrictions}
$\Gamma\vdash_{\mathbf{INF}^{\invertediota}} A$ iff $\Gamma\vdash_{\mathbf{INF}^{\invertediota}{'}}A$
\end{theorem}

\begin{proof}
From lemmas \ref{=genintroderivable}, \ref{=introderivable}, \ref{forallIexistsErestricted} and \ref{iotarestricted}. 
\end{proof}

\noindent We need to modify some definitions. $(=I^{nG})$ is often treated similarly to $(\exists E)$. The parameter of an application of $(=I^{nG})$ is assumed to be used solely for that application and to occur nowhere else in the deduction, and it also gives rise to segments: 

\begin{definition}
In definition \ref{segment}, insert `or $(=I^{nG})$' after $(\exists E)$.
\end{definition}

\noindent Definition \ref{maxformseg} stays the same. In Definition \ref{path}, paths continue with a discharged assumption from the premise of $(=I^{nG})$, but we won't have much occasion to consider this notion. 

Definition \ref{maxexistsidentitysegment} requires change and there is one more case to be considered: 

\begin{definition}
A maximal $=$-segment is (i) a segment of formulas $t=t$ such that one of them is the minor premise of $(=I^{nG})$ and the last is the premise of $(AD)$ or $(\invertediota E_3)$, or (ii) a segment (of formulas $a=t$) such that the first is discharged by $(=I^{nG})$ and the last is the premise of $(AD)$ or $(\invertediota E_3)$ with conclusion $\exists !t$. A maximal $\exists !$-segment is a segment that is concluded by $(AD)$ or $(\invertediota E_3)$ and major premies of an application of $(=I^{nG})$ with conclusion $t=t$.\end{definition}

\noindent If $\exists!a$ is concluded, this is not counted as maximal. Such segments are used in the proof that the rules for quantifiers and $\invertediota$ can be restricted. These are not detours and a formula different from the premise is concluded. 

The new reduction procedures are as follows:\bigskip 

\noindent I. Maximal $=$-segments (i): proceed from the rule that concludes the major premise to the rule applied to the maximal segment, removing all the rest. To illustrate the case where the last formula of the segment is the premise of $(AD)$ or $(\invertediota E_3)$: 

\begin{center}
\AxiomC{$\Sigma$}
\noLine
\UnaryInfC{$\exists !t$}
\AxiomC{$[a=t]^i$}
\noLine
\UnaryInfC{$\Pi$}
\noLine
\UnaryInfC{$t=t$}
\RightLabel{$_i$}
\BinaryInfC{$t=t$}
\UnaryInfC{$\exists !t$}
\DisplayProof\quad$\leadsto$\quad
\AxiomC{$\Sigma$}
\noLine
\UnaryInfC{$\exists!t$}
\DisplayProof
\end{center}

\noindent II. Maximal $=$-segments (ii): conclude the conclusion of $(AD)$ or $(\invertediota E_3)$ from whatever concludes the major premise of $(=I^{nG})$, leave everything else. To illustrate with a simple example: 

\begin{center}
\AxiomC{$\Pi$}
\noLine
\UnaryInfC{$\exists !t$}
\AxiomC{$[a=t]^i$}
\UnaryInfC{$\exists !t$}
\noLine
\UnaryInfC{$\Sigma$}
\noLine
\UnaryInfC{$C$}
\RightLabel{$_i$}
\BinaryInfC{$C$}
\DisplayProof\quad $\leadsto$ \quad
\AxiomC{$\Pi$}
\noLine
\UnaryInfC{$\exists !t$}
\AxiomC{$\Pi$}
\noLine
\UnaryInfC{$[\exists !t]$}
\noLine
\UnaryInfC{$\Sigma$}
\noLine
\UnaryInfC{$C$}
\RightLabel{$_i$}
\BinaryInfC{$C$}
\DisplayProof
\end{center}

\noindent The final application of $(=I^{nG})$ is required only if there are formulas in the assumption class of $a=t$ that are not premises of $(AD)$ or $(\invertediota E_3)$.\bigskip

\noindent III. Analogous to I: proceed from the rule that concludes the major premise to the rule applied to the maximal segment, removing all the rest. To illustrate the case where the first formula of the segment is the premise of $(AD)$ or $(\invertediota E_3)$: 

\begin{center}
\AxiomC{$\Sigma$}
\noLine
\UnaryInfC{$t=t$}
\UnaryInfC{$\exists !t$}
\AxiomC{$[a=t]^i$}
\noLine
\UnaryInfC{$\Pi$}
\noLine
\UnaryInfC{$t=t$}
\RightLabel{$_i$}
\BinaryInfC{$t=t$}
\DisplayProof\quad$\leadsto$\quad
\AxiomC{$\Sigma$}
\noLine
\UnaryInfC{$t=t$}
\DisplayProof
\end{center}

\noindent Should there be more than one application of $(=I^{nG})$ in the rules that give rise to the segment, pick the lowest one. 

\bigskip

\noindent Lemma \ref{removenewsegments} goes through with a slight modification. In case II, if more than one formula $a=t$ is discharged and $\Pi$ is not empty, the new reduction procedure increases the number of applications of rules in a deduction and multiplies any maximal $\exists!$-, $=$- or $(=E)$-segments in $\Pi$. 

\begin{lemma}\label{removenewsegmentsII}
Any deduction can be transformed into one without maximal $\exists!$-, $=$- and $(=E)$-segments. 
\end{lemma}

\begin{proof}
The procedures cannot introduce new maximal $(=E)$-segments. As before, the procedures may introduce new maximal $=$- and $\exists!$-segments or increase the length of existing ones, but as before in this case we remove them all together at once. To avoid multiplying maximal segments by procedure II, apply it to a maximal $=$-segment of this kind so that no other of that kind stands above it and remove all maximal $=$-segments of kind (i) and all maximal $(=E)$- and $\exists !$-segments that stand above it before applying the procedure. 
\end{proof}

\begin{theorem}
Deductions in $\mathbf{INF}^{\invertediota}{'}$ can be brought into normal form. 
\end{theorem}

\begin{proof}
By induction over the rank of deductions. 

The usual method for handling maximal segments works also when a formula concluded by $(=I^{nG})$ is major premise of an elimination rule: permute the application upwards. 

Vacuous applications of $(=E)$ are removed as before and assumed to be removed whenever they arise. 

With $(\forall E)$ and $(\exists I)$ restricted to atomic terms, the degree of any maximal formula affected by the replacement of parameters by terms in the reduction procedures stays the same. Thus the rank of deductions is reduced when the reduction procedures are applied according to the methods of Prawitz or Troelstra and Schwichtenberg. They can introduce new maximal $\exists!t$- or $=$-segments: these are removed by an appeal to Lemma \ref{removenewsegmentsII} immediately after carrying out the reduction procedure.

With $(\invertediota I)$, $(\invertediota E_1)$ and $(\invertediota E_2)$ restricted so that $t$ and $u$ are atomic, the reduction procedures for maximal formulas $\invertediota xF=t$ work as they should. Replacements of parameters by terms can no longer increase the degree of any maximal formulas. Any maximal formulas $F_u^x$ that are introduced by the procedure now have lower degree than the one removed, as $d(F_u^x)=d(\invertediota xF=t)-2)$. They can introduce vacuous applications of $(=E)$, which are removed as usual, and new maximal $\exists !$- and $=$--segments, which are removed by an appeal to Lemma \ref{removenewsegmentsII} immediately after the reduction.
\end{proof}

\subsection{Failure of the Subformula Property}
I shan't consider a modified subformula property for deductions in $\mathbf{INF}^{\invertediota}{'}$. There are too many exceptions. To give but one example, consider a derivation of `The $F$ is $G$', where $G$ is not $=$, arguably the more typical use of definite descriptions: 

 \begin{prooftree}
\AxiomC{$\Sigma_1$}
\noLine
\UnaryInfC{$\exists ! t$}
\AxiomC{$[a=t]^i$}
\noLine
\UnaryInfC{$\Xi$}
\noLine
\UnaryInfC{$F_a^x$}
\AxiomC{$[F_a^x]^j \ [\exists !a]^k$}
\noLine
\UnaryInfC{$\Pi$}
\noLine
\UnaryInfC{$a=t$}
%\LeftLabel{$(\invertediota I)$ \ }
\RightLabel{$_{i, j, k}$}
\TrinaryInfC{$\invertediota xF=t$}
\AxiomC{$\Sigma_2$}
\noLine
\UnaryInfC{$G_t^y$}
\BinaryInfC{$G_{\invertediota xF}^y$}
\end{prooftree}

\noindent $\invertediota xF=t$ would need to be exempt from the subformula property. 

There may be something systematic to the exceptions: they all involve $\exists !$ or $=$. But further investigation of this question must await another occasion.

\section{Normalisation for $\mathbf{CNF}$}
Here I shall be brief. Normalisation is a little easier, as there are no segments longer than $1$ to be considered, and paths are little simpler, too. But the rest stays more or less the same, and we have: 

\begin{theorem}
Deductions in $\mathbf{CNF}$ can be brought into normal form. 
\end{theorem}

\noindent Deductions in normal form in $\mathbf{CNF}$ have a version of the subformula property similar to what Prawitz finds in classical logic \citep[42]{prawitznaturaldeduction}: 

\begin{corollary}
Any exceptions to the subformula property in normal deductions in $\mathbf{CNF}$ occur between the first and the last formula of the M-part of paths or are assumptions $\neg A$ discharged  by $(\bot E_C)$ and formulas $\bot$ concluded from them. 
\end{corollary}

\section{Normalisation for $\mathbf{CNF}^{\invertediota}$}
$\mathbf{CNF}^{\invertediota}{'}$ arises from $\mathbf{CNF}^{\invertediota}$ by replacing $(=I^n)$ by $(=I^{nG})$, and restricting  $(\forall E)$ to atomic instantiating terms and $(\invertediota I)$, $(\invertediota E_1)$ and $(\invertediota E_2)$ to atomic $t$ and $u$. 

\begin{theorem}
$\Gamma\vdash_{\mathbf{CNF}^{\invertediota}} A$ iff $\Gamma\vdash_{\mathbf{CNF}^{\invertediota}{'}}A$
\end{theorem}

\begin{proof}As for theorem \ref{equivalencerestrictions}.\end{proof}

\noindent Recall that $(\bot E_C)$ with vacuous charges is treated like $(\bot E)$, i.e. its conclusions are restricted to prime conclusions (Lemma \ref{botrestriction}). If discharge is not vacuous, its conclusions cannot be so restricted. Atomic formulas containing $\invertediota$ terms must be admitted. We therefore need to consider further cases of maximal formulas, namely formulas concluded by $(\bot E_C)$ that are major premise of $(\invertediota E_1)$, $(\invertediota E_2)$, $(\invertediota E_3)$ or premise of $(AD)$. The last gives rise to an anomaly: the maximal formula may have degree $0$ if $t_i$ in the conclusion is atomic.  

The new reduction procedures follow the pattern of those given by St\aa lmarck in his normalisation proof for classical logic \citep[131ff]{stalmarckclassical}: they permute the applications of the elimination rules or $(AD)$ upwards and, assuming the negations of their conclusions, apply $(\bot E_C)$ to them instead. Here the latter only happens in two of the new cases, though. I shall adapt Andou's simplification of St\aa lmarck's method to the present case \citep{andounormalisation}.\footnote{\citet{mancosugalvanzach} give a detailed exposition of Andou's method, which is also from where I learnt about it.} In fact, due to the absence of $\exists$ and $\lor$ from $\mathbf{CNF}^{\invertediota}{'}$ and the ensuing possibility of restricting $(\bot E_C)$ to atomic formulas, the proof is simpler than Andou's. 

Following Andou, define a new kind of segment and when they are maximal: 

\begin{definition}\normalfont
A \emph{segment} is a sequence of formulas arising from applications of $(=I^{nG})$ as in Definition \ref{segment} or a sequence of formulas $A_1\ldots A_n$ such that $A_1$ is not the conclusion of $(\bot E_C)$, and for all $i$, $A_i$ is the minor premise of $(\rightarrow E)$ the major premise of which is discharged by $(\bot E_C)$, $A_{i+i}$ is the conclusion of that application of $(\bot E_C)$, and $A_n$ is not a minor premise of $(\rightarrow E)$ the major premise of which is discharged by $(\bot E_C)$. 
\end{definition}

\noindent Where there is need to distinguish the two, call the latter `$(\bot E_C)$-segments'. The length of segments is defined as always, their degree as in Definition \ref{degree}. 

\begin{definition}\normalfont\label{newmaxclassical}
Add the following at the end of Definition \ref{maxformseg} of maximal segment: `or a segment the last formula of which is the conclusion of $(\bot E_C)$ and the major premise of an elimination rule or the premise of $(AD)$'. 
\end{definition}

\noindent Crucial to Andou's method is the following: 

\begin{definition}\normalfont
An assumption discharged by $(\bot E_C)$  is \emph{regular} if it is the major premise of $(\rightarrow E)$. A proof is \emph{regular} if all assumptions discharged by $(\bot E_C)$ in it are regular. 
\end{definition}

\begin{lemma}
(a) Any proof can be transformed into a regular proof. (b) In a regular proof, any assumption discharged by $(\bot E_C)$ stands to the right of a formula on a $(\bot E_C)$-segment. 
\end{lemma}

\begin{proof}
(a) If $\neg A$ is discharged by $(\bot E_C)$ but not major premise of $(\rightarrow E)$, deduce it by $(\bot E_C)$ and discharge a regular assumption instead: 

\begin{prooftree}
\AxiomC{$[\neg A]^i$}
\AxiomC{$[A]^j$}
\BinaryInfC{$\bot$}
\RightLabel{$_j$}
\UnaryInfC{$\neg A$}
\end{prooftree}

\noindent where $i$ is the label of the original application of $(\bot E_C)$ and $j$ is fresh. (b) is immediate. 
\end{proof}

\noindent The effect of this lemma is that the minor premise of $(\rightarrow E)$ is always available in the upwards permutations of the elimination rules or $(AD)$. Remove all formulas in assumption class $i$ by concluding the conclusion of $(\rightarrow E)$ according to the pattern below and replace $(\bot E_C)$ accordingly or remove it altogether: 

\bigskip

\noindent 1. $(\bot E_C)$ followed by $(\invertediota E_1)$:
{\footnotesize
\begin{center}
\AxiomC{$[\neg \ \invertediota xF=t]^i$}
\AxiomC{$\Xi$}
\noLine
\UnaryInfC{$\invertediota xF=t$}
\LeftLabel{$_{(\rightarrow E)}$}
\BinaryInfC{$\bot$}
\noLine
\UnaryInfC{$\Pi$}
\noLine
\UnaryInfC{$\bot$}
\RightLabel{$_i$}
\LeftLabel{$_{(\bot E_C)}$}
\UnaryInfC{$\invertediota xF=t$}
\AxiomC{$\Sigma$}
\noLine
\UnaryInfC{$u=t$}
\LeftLabel{$_{(\invertediota E_1)}$}
\BinaryInfC{$F_u^x$}
\DisplayProof\quad$\leadsto$\quad
\AxiomC{$[\neg F_u^x]^j$}
\AxiomC{$\Xi$}
\noLine
\UnaryInfC{$\invertediota xF=t$}
\AxiomC{$\Sigma$}
\noLine
\UnaryInfC{$u=t$}
\LeftLabel{$_{(\invertediota E_1)}$}
\BinaryInfC{$F_u^x$}
\LeftLabel{$_{(\rightarrow E)}$}
\BinaryInfC{$\bot$}
\noLine
\UnaryInfC{$\Pi$}
\noLine
\UnaryInfC{$\bot$}
\RightLabel{$_j$}
\LeftLabel{$_{(\bot E_C)}$}
\UnaryInfC{$F_u^x$}
\DisplayProof
\end{center}
}
\noindent 2. $(\bot E_C)$ followed by $(\invertediota E_2)$:
{\footnotesize
\begin{center}
\def\defaultHypSeparation{\hskip .1in}
\AxiomC{$[\neg \invertediota xF=t]^i$}
\AxiomC{$\Xi$}
\noLine
\UnaryInfC{$\invertediota xF=t$}
\LeftLabel{$_{(\rightarrow E)}$}
\BinaryInfC{$\bot$}
\noLine
\UnaryInfC{$\Pi$}
\noLine
\UnaryInfC{$\bot$}
\RightLabel{$_i$}
\LeftLabel{$_{(\bot E_C)}$}
\UnaryInfC{$\invertediota xF=t$}
\AxiomC{$\Sigma_1$}
\noLine
\UnaryInfC{$F_u^x$}
\AxiomC{$\Sigma_2$}
\noLine
\UnaryInfC{$\exists !u$}
\LeftLabel{$_{(\invertediota E_2)}$}
\TrinaryInfC{$u=t$}
\DisplayProof \ $\leadsto$ \
\AxiomC{$[\neg u=t]^j$}
\AxiomC{$\Xi$}
\noLine
\UnaryInfC{$\invertediota xF=t$}
\AxiomC{$\Sigma_1$}
\noLine
\UnaryInfC{$F_u^x$}
\AxiomC{$\Sigma_2$}
\noLine
\UnaryInfC{$\exists !u$}
\LeftLabel{$_{(\invertediota E_2)}$}
\TrinaryInfC{$u=t$}
\LeftLabel{$_{(\rightarrow E)}$}
\BinaryInfC{$\bot$}
\noLine
\UnaryInfC{$\Pi$}
\noLine
\UnaryInfC{$\bot$}
\RightLabel{$_j$}
\LeftLabel{$_{(\bot E_C)}$}
\UnaryInfC{$u=t$}
\DisplayProof
\end{center}
}

\noindent 3. $(\bot E_C)$ followed by $(AD)$: 
{\footnotesize
\begin{center}
\AxiomC{$[\neg Rt_1\ldots t_n]^i$}
\AxiomC{$\Xi$}
\noLine
\UnaryInfC{$Rt_1\ldots t_n$}
\LeftLabel{$_{(\rightarrow E)}$}
\BinaryInfC{$\bot$}
\noLine
\UnaryInfC{$\Pi$}
\noLine
\UnaryInfC{$\bot$}
\RightLabel{$_i$}
\LeftLabel{$_{(\bot E_C)}$}
\UnaryInfC{$Rt_1\ldots Rt_n$}
\LeftLabel{$_{(AD)}$}
\UnaryInfC{$\exists !t_i$}
\DisplayProof\quad$\leadsto$\quad
\AxiomC{$\Xi$}
\noLine
\UnaryInfC{$Rt_1\ldots t_n$}
\LeftLabel{$_{(AD)}$}
\UnaryInfC{$\exists !t_i$}
\DisplayProof
\end{center}
}
\noindent $(\invertediota E_3)$ being a special case of $(AD)$, $(\bot E_C)$ followed by $(\invertediota E_3)$ is handled as in 3. 

\begin{theorem}
Deductions in $\mathbf{CNF}^{\invertediota}{'}$ can be brought into normal form.
\end{theorem}

\begin{proof}
First, transform the deduction into a regular deduction. Lemma \ref{removenewsegmentsII} goes through as before, so apply it next. Then proceed by an induction over the rank of deductions, defined as in Definition \ref{rank} adjusted so as to count the new maximal segments of Definition \ref{newmaxclassical}. Note that the highest degree of a maximal segment can now be $0$, namely if the formula on a maximal $(\bot E_C)$-segment is prime. Hence deductions may have rank $\langle 0, l\rangle$, where $l>0$, which cannot happen in standard classical logic. 

Maximal segments arising from $(=I^{nG})$ and from introducing and eliminating formulas as major premises are treated as usual. 

Maximal $(\bot E_C)$-segments are handled as follows. Due to the restriction of $(\bot E_C)$, their last formulas are atomic. All four procedures shorten or remove the maximal segment to which they are applied. 

In cases 1 and 2, any new maximal segments created by the reduction procedure have lower degree than the one shortened or removed, because due to the restriction on $(\invertediota I)$ $t$ and $u$ are atomic. Thus $d(F_u^x)=d(\invertediota xF=t)-2$ and $d(u=t)\leq d(\invertediota xF=t)-1$. 

In case 3, the following possibilities arise: 

(a) If $Rt_1\ldots t_n$ is prime (i.e. $R$ is a predicate letter), the degree of the maximal $(\bot E_C)$-segment is $0$, and cannot be lowered. But as the segment on which it is is shortened by $1$, this poses no problem. 

(b) If $Rt_1\ldots t_n$ is $\invertediota xF=t$ and it is concluded by $(\invertediota I)$ in $\Xi$, the procedure creates a maximal formula of the same degree as the one removed, so we remove it immediately by the third reduction procedure for maximal formulas $\invertediota xF=t$, leaving only the deduction of the premise $\exists !t$ of $(\invertediota I)$. 

(c) The possibilities removed with Lemma \ref{removenewsegmentsII} are: If $Rt_1\ldots t_n$ is concluded by $(=E)$ in $\Xi$, then the procedure introduces a maximal $(=E)$-segment. If $Rt_1\ldots t_n$ is $t=t$, then if it is concluded by $(=I^{nG})$ in $\Xi$ the procedure introduces a new maximal $=$-segment. It cannot introduce a new maximal $\exists !$-segment: this would require the conclusion $\exists !t$ of $(AD)$ to be major premise of $(=I^{nG})$, so there would have been a maximal $\exists !$-segment already, which were assumed to have been removed. 

The method by which Prawitz or Troelstra and Schwichtenberg choose a maximal segment to which to apply a permutative reduction procedure works for maximal $(\bot E_C)$-segments, too. It ensures that there are no longest maximal segments of highest degree in $\Xi$ or the $\Sigma$s, and so the rank of the deduction is lowered. 
\end{proof}

\section{Conclusion}
The normalisation theorems for systems of intuitionist and classical negative free logic without and with definite descriptions proved in this paper required considering new kinds of maximal formulas specific to negative free logic. The systems with definite descriptions required a formulation that avoids reduction procedures that involve arbitrarily complex terms being substituted for parameters. For the systems without definite descriptions, deductions in normal form have been shown to fulfil a restricted notion of subformula property. A system inspired by Ja\'skowski with an improved result has been proposed. The question remains whether there is an interesting notion of subformula property for the systems with definite descriptions. 

From the philosophical perspective, what is often considered to be a necessary condition for the meanings of expressions to be defined by rules of inference is thus fulfilled. A novelty of the present paper is that it opens up the prospects that the meanings of term-forming operators may also be so defined. Whether what has been established is also sufficient for proof-theoretic semantics is open to further discussion. Philosophical questions that arise have been touched upon. A satisfactory discussion of these issues must await another occasion. 

\bigskip

\noindent \textbf{Acknowledgements.} Thank you, as always, to Andrzej Indrzejczak, who encouraged me to work on this topic. This paper was improved substantially in response to the reports for this journal. I thank the referees for the exceptional refereeing and the time and effort they put into their thoughtful, encouraging and constructive criticism. 

\bigskip

\noindent\textbf{Funding.} The research in this paper was funded by the European Union (ERC, ExtenDD, project number: 101054714). Views and opinions expressed are however those of the author(s) only and do not necessarily reflect those of the European Union or the European Research Council. Neither the European Union nor the granting authority can be held responsible for them. 

\bigskip

\setlength{\bibsep}{0pt}
\bibliographystyle{chicago}
\bibliography{KurbisNormalNegFreeLogic}

\end{document}